\documentclass[pre,aps,twocolumn,superscriptaddress,showpacs]{revtex4-1}
\usepackage{amsmath,bm,graphicx}
\usepackage{subfigure}
\usepackage{amssymb}
\usepackage{amsfonts}
\usepackage{ifthen}
\newboolean{els}
\setboolean{els}{false}

\newcommand{\diffd}{\mbox{d}}
\newcommand{\sgn}{\mbox{sgn}}

\newcommand{\asinh}{\mbox{arsinh}\;}

\newcommand{\K}{\mathbf{k}}
\newcommand{\w}{\omega}
\numberwithin{equation}{section}

\begin{document}
\ifthenelse{\boolean{els}}
{
\begin{frontmatter}

\title{Externally forced triads of resonantly interacting waves: boundedness and integrability properties}

\author[WCCS]{Jamie Harris}
\author[UCD]{Miguel D. Bustamante}
\author[WMI,WCCS]{Colm Connaughton}

\address[WCCS]{Centre for Complexity Science, University of Warwick, Gibbet Hill Road, Coventry CV4 7AL, UK}
\address[UCD]{{School of Mathematical Sciences, University College Dublin, Belfield, Dublin 4, Ireland}}
\address[WMI]{Mathematics Institute, University of Warwick, Gibbet Hill Road, Coventry CV4 7AL, UK}
\begin{abstract}
We revisit the problem of a triad of resonantly interacting nonlinear waves
driven by an external force applied to the unstable mode of the triad. The
equations are Hamiltonian, and can be reduced to a dynamical system for 5 real variables with 2 conservation laws.
If the Hamiltonian, $H$, is zero we reduce this dynamical system to the motion
of a particle in a one-dimensional time-independent potential and prove that
the system is integrable. Explicit solutions are obtained for some
particular initial conditions. When explicit solution is not possible we present a novel numerical/analytical method for approximating the dynamics.
Furthermore we show analytically that when $H=0$ the motion is
generically bounded. That is to say the waves in the forced triad are bounded
in amplitude for all times for any initial condition with the single
exception of one special choice of initial condition for which the forcing is
in phase with the nonlinear oscillation of the triad. This means that the
energy in the forced triad generically remains finite for all time despite
the fact that there is no dissipation in the system. We provide a detailed
characterisation of the dependence of the period and maximum energy of the
system on the conserved quantities and forcing intensity.  When $H \neq 0$ we
reduce the problem to the motion of a particle in a one-dimensional
time-periodic potential. Poincar\'{e} sections of this system provide
strong evidence that the motion remains bounded when $H \neq 0$ and is
typically quasi-periodic although periodic orbits can certainly be found.
Throughout our analyses, the phases of the modes in the triad play a
crucial role in understanding the dynamics.
\end{abstract}

\end{frontmatter}
}
{
\title{Externally forced triads of resonantly interacting waves: boundedness and integrability properties}
\date{\today}

\author{Jamie Harris}
\email{jamie.harris@warwick.ac.uk}
\affiliation{Centre for Complexity Science, University of Warwick, Gibbet Hill Road, Coventry CV4 7AL, UK}

\author{Miguel D. Bustamante}
\email{miguel.bustamante@ucd.ie}
\affiliation{{School of Mathematical Sciences, University College Dublin, Belfield, Dublin 4, Ireland}}

\author{Colm Connaughton}
\email{connaughtonc@gmail.com}
\affiliation{Centre for Complexity Science, University of Warwick, Gibbet Hill Road, Coventry CV4 7AL, UK}
\affiliation{Mathematics Institute, University of Warwick, Gibbet Hill Road, Coventry CV4 7AL, UK}

\begin{abstract}
We revisit the problem of a triad of resonantly interacting nonlinear waves
driven by an external force applied to the unstable mode of the triad. The
equations are Hamiltonian, and can be reduced to a dynamical system for 5 real variables with 2 conservation laws.
If the Hamiltonian, $H$, is zero we reduce this dynamical system to the motion
of a particle in a one-dimensional time-independent potential and prove that
the system is integrable. Explicit solutions are obtained for some
particular initial conditions. When explicit solution is not possible we present a novel numerical/analytical method for approximating the dynamics.
Furthermore we show analytically that when $H=0$ the motion is
generically bounded. That is to say the waves in the forced triad are bounded
in amplitude for all times for any initial condition with the single
exception of one special choice of initial condition for which the forcing is
in phase with the nonlinear oscillation of the triad. This means that the
energy in the forced triad generically remains finite for all time despite
the fact that there is no dissipation in the system. We provide a detailed
characterisation of the dependence of the period and maximum energy of the
system on the conserved quantities and forcing intensity.  When $H \neq 0$ we
reduce the problem to the motion of a particle in a one-dimensional
time-periodic potential. Poincar\'{e} sections of this system provide
strong evidence that the motion remains bounded when $H \neq 0$ and is
typically quasi-periodic although periodic orbits can certainly be found.
Throughout our analyses, the phases of the modes in the triad play a
crucial role in understanding the dynamics.

\end{abstract}

\maketitle
}

\section{Introduction}
\label{sec-intro}

Resonant triads are the basic building blocks for understanding mode coupling in systems
of weakly interacting dispersive waves in which the leading order nonlinearity of the
underlying wave equation is quadratic in the wave amplitude, $\psi$. Physical examples of such
systems include capillary waves on fluid interfaces \cite{DKZ2003}, Rossby waves in 
geophysical fluid dynamics \cite{GIL1974,CNNQ2010}, mode coupling in nonlinear optics 
\cite{ABDP1962}, drift waves in magnetized plasmas \cite{HH1994}  and  internal waves in 
stratified fluids \cite{McCB1977, I2008}. For simplicity, let us suppose we can represent such wave 
fields in terms
of Fourier harmonics. Each linear mode can be labelled by its wave-vector, 
$\K \in \mathbb{R}^d$ (where $d$ is the physical dimension of the system),
and frequency, $\w_\K$. A resonant triad is a triplet of modes, $\K_1$, $\K_2$ and $\K_3$
which satisfy the resonance conditions:
\begin{eqnarray}
\nonumber \K_1 &=& \K_2 + \K_3\\
\label{eq-resonance} \w_{\K_1} &=& \w_{\K_2} + \w_{\K_3}.
\end{eqnarray}
In the limit of weakly nonlinear waves, energy is only exchanged efficiently between modes
which are members of resonant triads. For an idealised wave system in an unbounded
domain, the $\K$'s are continuous variables. A consequence of this fact is that a given $\K$ 
is typically a member of 
many resonant triads. The resonant triads therefore form a network in Fourier space which 
provides a conduit for energy transfer throughout the full set of Fourier modes. This idea 
of a network of resonant
modes carrying energy throughout the system is the principal basis for
the theory of wave turbulence \cite{ZLF92}.  Wave turbulence describes the statistical 
dynamics of an ensemble of weakly interacting dispersive waves in which many modes are 
excited by external forcing. In the usual wave turbulence scenario, which is relevant in
many applications, the forcing directly supplies energy only to a subset of these modes, 
typically those corresponding to large scales. Energy is then transferred to other modes,
in particular to small scale modes,
via resonant interactions in a process referred to as an energy cascade. The theoretical
description of such weakly nonlinear cascades is very well developed (see \cite{NR2011} for
a recent review).

For wave systems in bounded domains several interesting complications arise. The $\K$'s
are now discrete variables. In the simplest case, of a multiply periodic domain of size $L$,
the $\K$'s are of the form $\K_d\,\Delta k$ where $\Delta k = 2\,\pi/L$ and $\K_d \in
\mathbb{Z}^d$ is a vector of integers. Finding resonant triads in discrete systems is
therefore equivalent to finding integer-valued solutions of the resonance conditions,
Eq.~(\ref{eq-resonance}). In many applications, the dispersion relation, $\w_\K$, is
a power law, $\w_\K = k^\alpha$, often with a fractional exponent, $\alpha$. For such
dispersion relations, integer valued solutions  of the resonance conditions, if they exist
at all, are quite rare. Capillary waves ($\alpha=3/2$, $d=2$) are a well 
known example for which no discrete solutions of the resonance conditions are 
possible \cite{KAR1998}.
The sparsity of discrete solutions of the resonance conditions means that any particular $\K$
in a bounded wave system is typically only a member of relatively few resonant triads. 
Triads which share a common mode can be grouped together to form resonant clusters. While
many such clusters of varying size coexist in the Fourier space, no one cluster spans
the entire space and the resonant energy transfer network fragments into disconnected parts.
A full enumeration of the resonant clusters for Rossby waves is studied in \cite{LPPR2009}.
While very large clusters are typically few in number, there is usually a large
number of small clusters. In particular, isolated triads are quite common. An
isolated triad is a triad which shares no common mode with any other triads.
If the energy of the system is initially concentrated in the modes of a small
cluster, the simplest case being an isolated triad, then no resonant energy
cascade can take place. This gives rise to the phenomenon of 
``frozen turbulence'' \cite{PZ2000} in which energy remains localised 
in relatively few modes rather than cascading throughout the system. In
practice, the level of nonlinearity is always finite leading to nonlinear
broadening or detuning of resonances. This finite amplitude effect then allows 
energy to be exchanged between triads which are not exactly resonant so that a 
cascade may proceed even in the absence of a large resonant cluster spanning
the system \cite{CNP01}. Such quasi-resonant interactions are less efficient at transferring
energy in the sense that the timescale for energy transfer is longer compared
with resonant interactions. For this reason, the exactly resonant clusters, 
in particular the isolated triads, remain the building blocks for 
understanding energy transfer in the system when the level of nonlinearity is
finite but small.

After suitable rescalings of the wave amplitudes, the equations for an 
isolated triad can always be brought to the canonical form:
\begin{equation}
\label{eq-unforcedTriad}
\left\{
\begin{array}{rcl}
\dot{B}_1& =& Z B_2^*B_3,\\
\dot{B}_2& =& Z B_1^*B_3,\\
\dot{B}_3& =& -Z B_1B_2,
\end{array}
\right.
\end{equation}
where $B_1$, $B_2$ and $B_3$ are the rescaled complex amplitudes of the modes
constituting the triad. This is a Hamiltonian system with Hamiltonian
\begin{equation*}
H = 2 Z \mbox{Im}\{ B_1B_2B_3^*\}.
\end{equation*}
With this Hamiltonian, Eqs.~(\ref{eq-unforcedTriad}) are equivalent to 
Hamilton's  equations:
\begin{displaymath}
i\dot{B}_j = \frac{\partial H}{\partial B_j^*},\; j = 1,2,3
\end{displaymath}
along with their complex conjugates. $H$ is obviously conserved by these
dynamics.

Eqs.~(\ref{eq-unforcedTriad}) have been extensively studied for many
years since early work in nonlinear optics \cite{ABDP1962}, plasmas, atmospheric dynamics  \cite{LHG1967}  and 
electronics led
to an appreciation of the importance of mode coupling in nonlinear systems. 
They describe the nonlinear saturation of the so-called decay instability
which manifests itself in a variety of wave systems with quadratic nonlinearity
including capillary waves \cite{DKZ2003}, Rossby waves \cite{GIL1974,CNNQ2010}
and drift waves \cite{HH1994}. The decay instability, in its linear stage,
describes a process whereby modes $B_1$ and $B_2$, which initially contain
little energy, grow in amplitude at the expense of mode $B_3$ which initially
contains most of the energy. For this reason, $B_3$ is referred to as the
unstable mode of the triad. 
More recently, Eqs.~(\ref{eq-unforcedTriad}) have been studied in a mechanical 
context to describe the
dynamics of a spring pendulum \cite{LYN2003,LH2004}. Moreover there has been 
considerable recent interest in using these equations as models of resonant 
Rossby wave interactions to attempt to explain the observed periods of various
intra-seasonal oscillations in the Earth's atmosphere \cite{KL2007}.

It has been known for a long time that Eqs.~(\ref{eq-unforcedTriad}) form
an integrable system (for a discussion see \cite{BK2009B}). Analytic formulae
for the amplitudes of the modes in terms of elliptic functions have been 
known at least since the early 1960's \cite{ABDP1962}. Recently, corresponding 
explicit formulae for the phases, as encoded by the dynamics of the so-called 
dynamical phase, were obtained in \cite{BK2009}. This work means that 
everything is effectively known analytically about the dynamics of an isolated 
triad.

Given the role played by resonant triads in the theory of wave turbulence
discussed above, it is somewhat surprising that almost nothing is known 
about the dynamics of an isolated triad in the presence of forcing. We aim to
address this in the present paper. We consider the simplest possible forcing:
an additive forcing of the unstable mode. We force the unstable mode because
this will allow efficient transfer of energy into the triad since the unstable
modes tends to transfer energy to the other two members of the triad whereas
energy in the stable modes tends to remain there. Besides its ubiquity in
the study of wave turbulence at both the numerical and analytic level, the
addition of forcing to the equations of motion has direct physical meaning
in certain applications. In the case of Rossby waves, for example, an 
additive forcing term in Eqs.~(\ref{eq-unforcedTriad}) describes orographic
forcing of Rossby waves by an idealised periodic topography \cite{PED1987}.
For the swinging spring it would correspond to a periodic oscillation of the
point of support. A numerical study \cite{LYN2009} of the forced triad equations suggested
that the amplitudes of the modes remain bounded in the presence of forcing,
something which is not obvious \emph{a priori}.

In this paper we provide a detailed study of the properties of an isolated
triad in the presence of forcing. The paper is laid out as follows. 
We begin in Sec.\ref{sec-forcedTriad} by writing the equations of the
forced triad in the general case and studying their conservation laws.
We then specialise in Sec.~\ref{sec-HZero} to studying the case $H=0$. 
We prove that the system is generically bounded when $H=0$ meaning that it is 
bounded for any choice of initial condition with the exception of a 
single special initial condition for which the system is unbounded (see Sec. 
\ref{sec-unbounded}).  We obtain an explicit solution in 
Sec.\ref{sec-explicitSolution} of the forced triad evolution equations
for a particular choice of initial conditions. An explicit solution of the evolution equations in terms of elliptic functions does
not seem possible in general, even in the integrable case $H=0$.
Instead, in Sec. ~\ref{sec-parametricStudy} we provide formulae for the period 
and maximum amplitude of the solution which are valid in general. We 
perform a parametric study of the period and maximum amplitude of the triad as 
key parameters are varied. 
Then in Sec.~\ref{sec-approximateSolutions}, we introduce a novel and 
accurate scheme for approximating the $H=0$ dynamics and compare it against 
numerical solutions of the original system for a variety of initial conditions.
Finally, in Sec. ~\ref{sec-HNonZero}, we turn our
attention to the case $H\neq 0$. We find that the dynamics can be reduced
to the one-dimensional motion of a particle in a time-periodic potential.
Numerical Poincar\'{e} sections of the resulting dynamical system suggest that the
dynamics remain bounded when $H \neq 0$ and the motion is generally quasi-periodic
although periodic orbits can also be identified. The article then ends with a 
brief summary and conclusions.
\section{General Forced Triad}
\label{sec-forcedTriad}

Let $B_1$, $B_2$ and $B_3$ be complex functions of time. We assume they satisfy a system of evolution equations derived from the real Hamiltonian
\begin{equation*}
H = 2 Z \mbox{Im}\{ B_1B_2B_3^*\} - 2 {F\,}\mbox{Im}\{iB_3\},
\end{equation*}
where ${F\,}$ and $Z$ are constant real parameters. The canonical equations of motion for this system are given by
\begin{displaymath}
i\dot{B}_j = \frac{\partial H}{\partial B_j^*},\; j = 1,2,3
\end{displaymath}
along with their complex conjugates. The equations of motion are, therefore,
\begin{equation}
\label{equations_of_motion}
\left\{
\begin{aligned}
\dot{B}_1& = Z B_2^*B_3,\\
\dot{B}_2& = Z B_1^*B_3,\\
\dot{B}_3& = -Z B_1B_2 + i{F\,},
\end{aligned}
\right.
\end{equation}
together with their complex conjugates. In this setting, the term `$i{F\,}$' is originated from a forcing of the so-called `unstable' mode represented by $B_3.$

\subsection{Amplitude-Phase Representation}

We write  $ B_j = C_j \mbox{e}^{i\varphi_j},\; j = 1,2,3, $
with $C_j\in\mathbb{R}$ and $\varphi_j\in\mathbb{R}$ to obtain
\begin{displaymath}
\dot{B}_j = \left( \dot{C}_j + i C_j \dot{\varphi}_j\right)\mbox{e}^{i\varphi_j}.
\end{displaymath}
Under this new representation, the equations of motion (\ref{equations_of_motion}) become
\begin{equation}
\label{p-a_equations_of_motion}
\left\{
\begin{aligned}
\dot{C}_1& = Z C_2 C_3 \cos\varphi,\\
\dot{C}_2& = Z C_1 C_3 \cos\varphi,\\
\dot{C}_3& =-Z C_1 C_2 \cos\varphi + {F\,}\sin\varphi_3,\\
\dot{\varphi}_1& = -Z \frac{C_2 C_3}{C_1} \sin\varphi,\\
\dot{\varphi}_2& = -Z \frac{C_1 C_3}{C_2} \sin\varphi,\\
\dot{\varphi}_3& = -Z \frac{C_1 C_2}{C_3} \sin\varphi+\frac{{F\,}}{C_3}\cos\varphi_3,\
\end{aligned}
\right.
\end{equation}
where we have introduced the dynamical phase $\varphi = \varphi_1+\varphi_2-\varphi_3$.

\subsection{Conservation Laws of General Forced Triad and reductions}

The unforced triad (i.e., the case $F=0$) is known to possess two independent conservation laws that are quadratic in the amplitudes $C_j$, and one conservation law which is cubic in the amplitudes. In contrast, the forced triad possesses one quadratic conservation law and one cubic conservation law. These are
\begin{eqnarray}
\nonumber
J &=& C_1^2 - C_2^2,\\
\label{Hamiltonian}
H &=& 2 C_3\left[ ZC_1C_2\sin\varphi - {F\,}\cos\varphi_3\right].
\end{eqnarray}
Here, $H$ is nothing but the Hamiltonian of the original system (\ref{equations_of_motion}), written in terms of the new variables.

Another important property of the unforced triad is that all the phases $\varphi_1$, $\varphi_2$ and $\varphi_3$ were `slave' variables, i.e., they were obtainable by quadratures once the solutions for $C_1$, $C_2$, $C_3$ and $\varphi$ were known. In contrast, in the forced triad only $\varphi_1$ and $\varphi_2$ are `slave' variables, and the five real variables $C_1$, $C_2$, $C_3$, $\varphi$ and $\varphi_3$ form the reduced system
\begin{equation}
\label{reduced_system}
\left\{
\begin{aligned}
\dot{C}_1& = Z C_2 C_3 \cos\varphi,\\
\dot{C}_2& = Z C_1 C_3 \cos\varphi,\\
\dot{C}_3& = -Z C_1 C_2 \cos\varphi + {F\,}\sin\varphi_3,\\
\dot{\varphi}_3& = -\frac{H}{2C_3^2},\\
\dot{\varphi}& = -Z C_1C_2C_3\sin\varphi\left(\frac{1}{C_1^2} + \frac{1}{C_2^2}\right) + \frac{H}{2C_3^2},
\end{aligned}
\right.
\end{equation}
where $H$ is defined as in equation (\ref{Hamiltonian}). Numerical simulations of system (\ref{reduced_system}) can be interpreted as either using $H$ as a constant or by explicitly using $H$ as defined in equation (\ref{Hamiltonian}). The accuracy might be sensitive to this choice. 

System (\ref{reduced_system}) along with the two conservation laws, $J$ and $H$, is effectively a three-dimensional system and as such it might not be integrable. We notice, however, that the system is volume-preserving, with Jacobi last multiplier
\begin{equation*}
\rho = 
\begin{cases}
C_1C_2,&\mbox{if $H$ = constant},\\
C_1C_2C_3,&\mbox{if $H$ is defined as in Eq.(\ref{Hamiltonian})}.
\end{cases}
\end{equation*}

\noindent \textbf{Remark:} a Jacobi last multiplier, also known as a standard Liouville volume density, is a scalar function $\rho(\mathbf{x})$ of the dependent variables $x^a\,,\quad a = 1, \ldots, N,$ defined by the equation $\nabla \cdot (\rho \, \mathbf{V}(\mathbf{x})) = 0\,,$ equation to be valid for all values of $\mathbf{x},$ where $\mathbf{V}$ is the right-hand-side of the evolution equations $\dot{\mathbf{x}} = \mathbf{V}(\mathbf{x}).$ See \cite{H1992,BH2003,BK2011}, \cite{NL2008} and references therein.

\subsection{Integrability of the case $H=0$}

We consider initial conditions for system (\ref{reduced_system}) such that $H=0.$ We obtain
\begin{equation}
\label{reduced_system_H_is_zero}
\left\{
\begin{aligned}
\dot{C}_1&= Z C_2 C_3 \cos\varphi,\\
\dot{C}_2&= Z C_1 C_3 \cos\varphi,\\
\dot{C}_3&= -Z C_1 C_2 \cos\varphi + {F\,}\sin\varphi_3,\\
\dot{\varphi}_3&= 0,\\
\dot{\varphi}&=-Z C_1C_2C_3\sin\varphi\left(\frac{1}{C_1^2} + \frac{1}{C_2^2}\right).
\end{aligned}
\right.
\end{equation}
We notice that the case $C_3=0$ is trivially integrable. Assuming that $C_3\not\equiv 0$, we obtain $H=0\Leftrightarrow ZC_1C_2\sin\varphi = {F\,}\cos\varphi_3$. But, according to the dynamics (\ref{reduced_system_H_is_zero}), $\varphi_3$ is constant, so we derive the new conservation law
\begin{equation}
\label{eq:new_const}
C_1C_2\sin\varphi = \frac{{F\,}\cos\varphi_3}{Z}.
\end{equation}
This new constant, together with $J=C_1^2 - C_2^2$, helps us reduce the system to an effectively two-dimensional volume-preserving and therefore integrable system:
\begin{equation}
\label{reduced_system_H_is_zero-new_cons}
\left\{
\begin{aligned}
\dot{C}_1& = Z C_2 C_3 \cos\varphi,\\
\dot{C}_2& = Z C_1 C_3 \cos\varphi,\\
\dot{C}_3& = -Z C_1 C_2 \cos\varphi + {F\,}\sin\varphi_3,\\
\dot{\varphi}& = -{F\,}C_3\cos\varphi_3\left(\frac{1}{C_1^2} + \frac{1}{C_2^2}\right),
\end{aligned}
\right.
\end{equation}
with $\varphi_3$ being constant, together with the conservation laws
\begin{eqnarray}
\label{new_conservation_law}
C_1C_2\sin\varphi &=& \frac{{F\,}\cos\varphi_3}{Z},\\
\label{J_cons_law}
C_1^2 - C_2^2 &=& J,
\end{eqnarray}
and Jacobi last multiplier given by
\begin{equation*}
\rho = 
\begin{cases}
C_1C_2,& \mbox{if } F\cos\varphi_3 = ZC_1C_2\sin\varphi\\
&\mbox{is used in Eq.(\ref{reduced_system_H_is_zero-new_cons})},\\
1,&\mbox{if } F\cos\varphi_3 = \mbox{constant} \\
&\mbox{is used  in Eq.(\ref{reduced_system_H_is_zero-new_cons})}.
\end{cases}
\end{equation*}

\section{Generic boundedness of the integrable case $H=0$}
\label{sec-HZero}
The goal within this section is to establish generic boundedness of the solutions to the integrable system (\ref{reduced_system_H_is_zero-new_cons})--(\ref{J_cons_law}). By boundedness we mean that the `energy' $E(t)$ of the system is bounded for all times. The energy is a positive-definite quadratic function of the amplitudes $C_j(t)$:
\begin{equation}
\label{def_energy}
 E(t) = \frac{C_1(t)^2+C_2(t)^2}{2} + C_3(t)^2.
\end{equation}
This quantity is obtained from the kinetic energy of the original wave system and should not be confused with the Hamiltonian $H,$ which is a cubic function of the amplitudes and equal to zero in this case. 

The time derivative of the energy is obtained using equations (\ref{reduced_system_H_is_zero-new_cons}):
\begin{eqnarray}
\nonumber
 \dot E(t) &=& C_1(t) \dot C_1(t) + C_2(t) \dot C_2(t) + 2 \, C_3(t) \dot C_3(t)\\
\label{dot_energy}
           &=& 2\,{F\,} \sin \varphi_3\,C_3(t)\,.
\end{eqnarray}
This result suggests that we use a `local rescaled time' variable $\tau(t)$ satisfying
\begin{equation}
\label{relationship-tau_and_C_3}
\frac{\mbox{d}\tau}{\mbox{d}t} = C_3(t)\,,\quad \tau(0) = 0\,.
\end{equation}
This time transformation is well defined as long as $C_3(t)$ is not zero. Integrating equation (\ref{dot_energy}) we obtain
\begin{equation}
\label{eq_E}
E(t) = E(0) + 2\, \tau \,{F\,} \sin \varphi_3\,, 
\end{equation}
where $\tau=\tau(t).$ We see that proving boundedness of energy is equivalent to proving boundedness of $\tau(t).$

\subsection{Establishing boundedness for $J \neq 0$}
\label{subsec:boundedness}
The conservation law $J = C_1^2-C_2^2$ naturally suggests a parameterisation of the form:
\begin{equation*}
\left\{
\begin{aligned}
C_1(t)&=\sqrt{J}\;\sgn(C_1(0))\cosh\omega(\tau),\\
C_2(t)&=\sqrt{J}\sinh\omega(\tau),
\end{aligned}
\right.
\end{equation*}
in the case $J>0$ and
\begin{equation*}
\left\{
\begin{aligned}
C_1(t)&=\sqrt{-J}\sinh\omega(\tau),\\
C_2(t)&=\sqrt{-J}\;\sgn(C_2(0))\cosh\omega(\tau),
\end{aligned}
\right.
\end{equation*}
when $J<0.$ Substituting this parameterised solution into equations (\ref{reduced_system_H_is_zero-new_cons}) and using equations (\ref{new_conservation_law}), (\ref{J_cons_law}) and (\ref{relationship-tau_and_C_3}), we deduce that the function $\omega(\tau)$ must satisfy, regardless of whether $J$ is positive or not,
\begin{equation*}
\cosh 2\omega(\tau) = \frac{2}{|J|}\,R \cosh\left(2Z\tau + {\Delta} \right),
\end{equation*}
where $R > 0$ and $\Delta \in \mathbb{R}$ are constants, related to the initial conditions as follows:
\begin{eqnarray}
\label{def_R}
R &=& \sqrt{\left(\frac{J}{2}\right)^2+\frac{{F\,}^2\cos^2\varphi_3}{Z^2}}\,,\\
\label{def_Delta}
R\,\sinh {\Delta} &=&  C_1(0)\,C_2(0)\,\cos \varphi(0)\,  .
\end{eqnarray}

Substituting the solution for $\omega(\tau)$ into the parameterised forms for $C_1(t)$ and $C_2(t)$ gives the result valid for $J\neq 0:$ 
\begin{equation}
\label{C_1(tau)_C_2(tau)_squared}
\left\{
\begin{aligned}
C_1(t)^2 &=  R\,\cosh\left(2Z\tau +{\Delta}\right) +\frac{J}{2},\\
C_2(t)^2 &=  R\,\cosh\left(2Z\tau + {\Delta}\right)\,-\,\frac{J}{2}.
\end{aligned}
\right.
\end{equation}
The solution for the remaining amplitude $C_3(t)$ is found by using equations (\ref{def_energy}) and (\ref{eq_E}). We get
\begin{eqnarray}
\label{C_3_squared-tau}
C_3(t)^2 =  E(0) + 2 {F}\tau \sin\varphi_3 -{R}\cosh\left(2Z\tau+\Delta\right),
\end{eqnarray}
and using equation (\ref{relationship-tau_and_C_3}) we derive the following equation for $\tau(t)$:
\begin{equation}
\label{differential_eqn_tau}
\left[\frac{\mbox{d}\tau}{\mbox{d}t}\right]^2 + R \cosh\left(2Z\tau+\Delta\right) -2 F\tau\sin\varphi_3 = E(0).
\end{equation}
This is precisely the equation for a one-dimensional particle with position $\tau(t)$, total energy $E(0),$ kinetic energy $\left[\frac{\mbox{d}\tau}{\mbox{d}t}\right]^2$ and potential energy
\begin{equation}
\label{potential-bounded}
V(\tau) = R \cosh\left(2Z\tau+\Delta\right) -2 F\tau\sin\varphi_3\,.
\end{equation}
From the fact that $R>0$ it follows that this potential is in fact attractive: $V''(\tau)>0$ for all $\tau\in\mathbb{R}$, and  $\lim_{\tau\rightarrow\pm\infty}V(\tau) = \infty.$ It follows that the particle must oscillate about a unique minimum situated at the point
\begin{equation*}
\tau_* = -\frac{\Delta}{2Z} + \frac{1}{2Z}\sinh^{-1}\left(\frac{F\sin\varphi_3}{Z R}\right),
\end{equation*}
with the motion of $\tau(t)$ being bounded between the two turning points $\tau_{\min} < \tau_{\max}$ defined by the solutions of 
\begin{equation*}
V(\tau_{\min}) =  V(\tau_{\max}) = E(0).
\end{equation*}
Since $\tau(t)$ is bounded, we conclude from equations (\ref{C_1(tau)_C_2(tau)_squared}) and (\ref{C_3_squared-tau}) that the amplitudes of each mode must also be bounded.\\

\noindent \textbf{Evolution of the dynamical phase $\varphi(t).$} Equation (\ref{new_conservation_law}) could at first sight be used to solve for the dynamical phase $\varphi(t),$ however this is not practical because the dynamical phase generically oscillates about $\pi/2$ (modulo $n \pi$) and therefore the inverse sine is multivalued. It is better to integrate directly equation (\ref{reduced_system_H_is_zero-new_cons}) for $\varphi$ using the auxiliary variable $\tau(t).$ The result is: if the initial condition $\varphi(0)$ is in the open interval $(n \pi, (n+1)\pi),$ for some $n \in \mathbb{Z},$ then $\varphi(t)$ remains in that interval and
\begin{equation}
\label{dynamical_phase_H=0}
 \varphi(t) = \cot^{-1} \frac{Z R \sinh(2Z \tau + \Delta)}{F \cos \varphi_3}\,,
\end{equation}
where the branch is properly selected. In the special case when the constant $\varphi_3$ is equal to $(m+1/2) \pi \,,\,\, m \in \mathbb{Z},$ then $\varphi$ is also constant and equal to $\tilde{m} \pi \,,\,\,\tilde{m} \in \mathbb{Z}\,.$ \\

\subsection{Formulae for the turning points in terms of new special functions}

The turning points $\tau_{\min/\max}$ of the evolution equation (\ref{differential_eqn_tau}) are  defined by the solutions of the equation $V(\tau) = E(0):$ 
\begin{equation}
\label{turning_points}
R \cosh\left(2Z\tau+\Delta\right) -2 F\tau\sin\varphi_3 = E(0). 
\end{equation}
For physically admissible values of the parameters $Z, F$ and initial conditions $R, \varphi_3, E(0),$ this equation has either two real solutions or only one real solution. When there is only one real solution, the physical system is ``dynamically frozen'' at $\tau = \mbox{constant}.$ In order to study a less trivial nonlinear dynamics, it is better to restrict the parameter space so that there are two real solutions, denoted by $\tau_{\min}, \tau_{\max},$ with $\tau_{\min}<\tau_{\max}.$ 

Let us study the structure of these solutions as functions of the parameters. Notice that if we replaced the hyperbolic cosine in equation (\ref{turning_points}) by an exponential, the solution for $\tau_{\min/\max}$ would be given in terms of two branches of the product log function $W_k(x)$, a function of one variable. In contrast, in our case such reduction does not get too far: the solution is given in terms of a function of two variables. Defining the function $\xi(x,y)$ by the implicit equation
\begin{equation}
 \label{xi_function}
\cosh \xi = x \, \xi + y\,,
\end{equation}
we look for real solutions $\xi \in \mathbb{R}.$  Graphically, we want to find the intersections between two graphs, $\cosh \xi$ and $x \xi + y,$ as function of $x$ and $y.$ We see that tangent intersection occurs if in addition to the above equation, $\sinh \xi = x$ holds. Therefore, a relation between the arguments is found in the tangent case:
$y = \sqrt{1+x^2} - x \,\sinh^{-1}x\,.$ Therefore, the condition (of physical origin) to have two solutions is a restriction on the parameter space:
\begin{equation}
 \label{xi_restriction}
y > \sqrt{1+x^2} - x \,\sinh^{-1}x\,, \quad x \in \mathbb{R}\,.
\end{equation}
With this condition, there will be two real solutions of equation (\ref{xi_function}), denoted
$$\xi_{-1}(x,y) < \xi_0(x,y)\,.$$
With these functions defined, the turning points $\tau_{\min}, \tau_{\max}$ can be obtained via proper rescalings:
\begin{equation}
\label{turning_points:xi}
2 Z \, \tau_{\min/\max} + \Delta = \xi_k \left(\frac{F \sin \varphi_3}{R\,Z}, \frac{E(0)}{R} - \frac{F \sin \varphi_3 \, \Delta}{R\,Z}\right)\,,
\end{equation}
where $k=0$ or $-1$. In practice, we resort to a numerical algorithm to compute these solutions.

Condition (\ref{xi_restriction}) is automatically satisfied for any choice of real initial amplitudes and phases.

\subsection{An explicit example of bounded motion: $\sin\varphi_3 = 0, J\neq 0$}

\label{sec-explicitSolution}

Continuing with the case $H=0,$ consider now a choice of initial conditions such that $\varphi_3\in\{0,\pm\pi\}$ and $J\neq0.$ The differential equation for $\tau(t)$, equation (\ref{differential_eqn_tau}), now reads
\begin{equation}
\label{differential_eqn_tau-explicit}
\left[\frac{\mbox{d}\tau}{\mbox{d}t}\right]^2 = E(0) - R\,\cosh\left(2Z\tau+\Delta\right) .
\end{equation}
The particle interpretation is simple, given the fact that $E(0) > R$ by virtue of the condition $H=0.$ It turns out that the potential $V(\tau)$, apart from a shift, is a symmetric well and therefore the motion should be bounded. This equation can be explicitly integrated to give an analytical solution for $\tau(t)$ in terms of Jacobi elliptic functions, with modulus $m = \frac{E(0)-R}{E(0)+R} \in (0,1)\,.$ The procedure is standard but tedious, and is omitted here. The result can be checked directly by substitution into equation (\ref{differential_eqn_tau-explicit}). We obtain the explicit solution
$$2\,Z\,\tau(t) + \Delta = \ln \left({\frac{ \scriptstyle \left[ \mathrm{dn}\left(\left.\frac{2\,K(m)}{T}\,t + u\right|m\right) - \sqrt{m}\,\mathrm{cn}\left(\left.\frac{2\,K(m)}{T}\,t + u\right|m\right)\right]^2}{1-m}}\right)\,, $$
where $K(m)$ is the complete elliptic integral of the first kind, the period $T$ of the oscillations is defined by
\begin{equation*}
T = \frac{2\,K(m)}{Z\,\sqrt{E(0)+R}},
\end{equation*}
and the shift $u$ is defined in terms of the incomplete elliptic integral of the first kind by 
\begin{equation*}
u = F\left(\left.\frac{C_3(0)}{\sqrt{E(0)-R}}\right|m\right).
\end{equation*}

\begin{figure}
\begin{center}
\includegraphics[width=70mm]{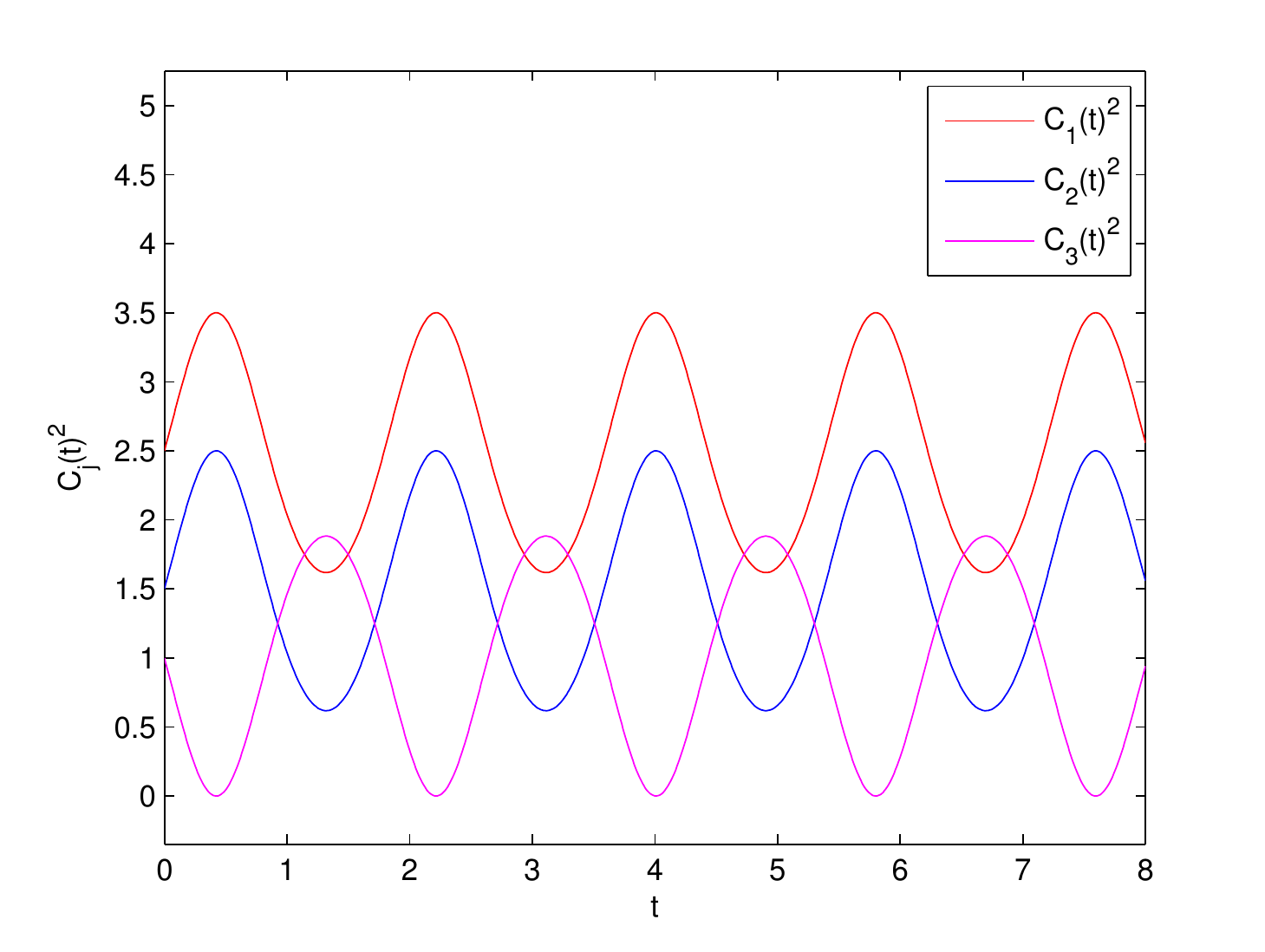}
\includegraphics[width=70mm]{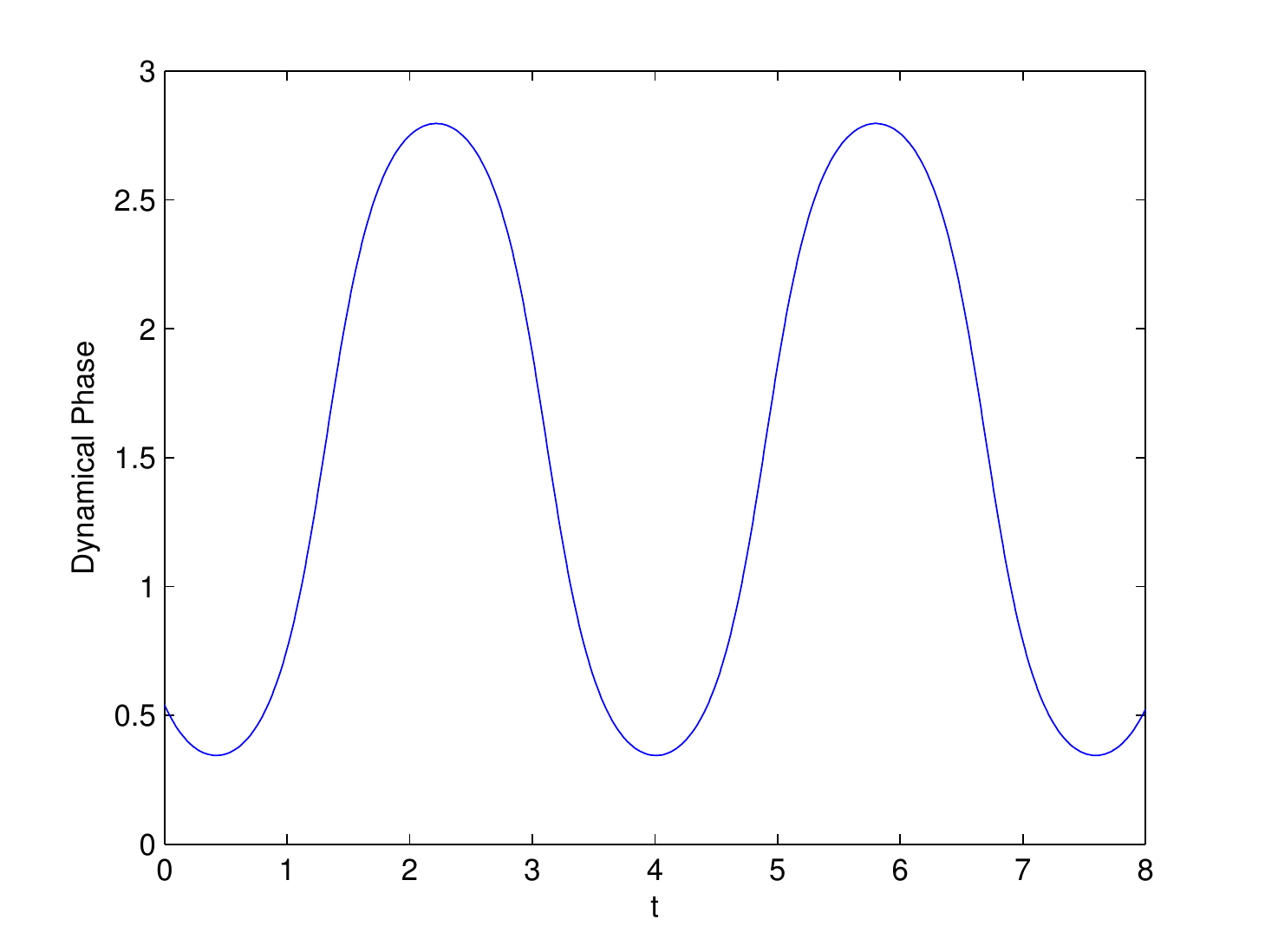}
\caption{\label{fig_explicit} Solutions for the square amplitudes (top panel) and dynamical phase (bottom panel) in the explicitly integrable case $H=0$ and $\varphi_3 = 0$. Choices for parameters and initial conditions: $F=1$, $Z=1,$ $E(0)=3$, $J=1$, and $\varphi(0)$ is obtained from equation (\ref{eq:dyn_phase_expl}).}
\end{center}
\end{figure}

To see that the above solution for $\tau(t)$ is indeed oscillatory, we compute its time derivative: 
\begin{equation*}
\frac{\diffd\tau}{\diffd t} = \sqrt{E(0)-R}\;\mathrm{sn}\left(\left.\frac{2\,K(m)}{T}\,t + u\right|m\right).
\end{equation*}
With this result, we readily obtain the squares of the amplitudes $C_1(t), C_2(t), C_3(t):$
\begin{equation*}
\left\{
\begin{aligned}
C_1(t)^2 &= {\scriptstyle E(0)+\frac{J}{2}}-{\scriptstyle\left(E(0)-R\right)}\;\mathrm{sn}^2\left(\left.\frac{2\,K(m)}{T}\,t + u\right|m\right),\\
C_2(t)^2 &={\scriptstyle E(0)-\frac{J}{2}}-{\scriptstyle\left(E(0)-R\right)}\;\mathrm{sn}^2\left(\left.\frac{2\,K(m)}{T}\,t + u\right|m\right),\\
C_3(t)^2 &={\scriptstyle \left(E(0)-R\right)}\;\mathrm{sn}^2\left(\left.\frac{2\,K(m)}{T}\,t + u\right|m\right).
\end{aligned}
\right.
\end{equation*}
We plot these squares in figure \ref{fig_explicit}, top panel. Finally, from equation (\ref{dynamical_phase_H=0}) we obtain the dynamical phase: 
\begin{equation}
 \label{eq:dyn_phase_expl}
\varphi(t) = -\cot^{-1} \left[{\textstyle \frac{Z \,R \,m^{1/2} }{F \,(1-m)\, \cos \varphi_3} \mathrm{cn\,dn}\left(\left.\frac{2\,K(m)}{T}\,t + u\right|m\right)}\right]\,,
\end{equation}
where we have defined $\mbox{cn dn}(x|m)\equiv\mbox{cn}(x|m)\,\mbox{dn}(x|m)$. Depending on the initial value $\varphi(0),$ the dynamical phase remains bounded and oscillates about the value $(n+1/2)\pi\,,$ for some $n \in \mathbb{Z}.$ The corresponding plot of the solution for the dynamical phase is shown in figure \ref{fig_explicit}, bottom panel.

\subsection{The limit $J \to 0$: ``almost always'' bounded}

Always in the case $H=0,$ we have established boundedness of the motion for the case $J\neq0$ in Section \ref{subsec:boundedness}. Using the equations from that Section, it is possible to establish boundedness in the case $J=0, \cos \varphi_3 \neq0,$ by taking the limit of the $J\neq0$ equations as $J\to0.$ For example, the auxiliary function $\omega(\tau)$ has a singularity in the limit:
$$\omega(\tau) \approx \frac{1}{2} \log \left[\frac{4 |F \cos \varphi_3|}{|J \,Z|} \cosh(2\,Z\,\tau+\Delta)\right]\,,$$
but the physical quantities such as amplitudes and phases behave well: in particular, equations (\ref{C_1(tau)_C_2(tau)_squared}) and (\ref{dynamical_phase_H=0}) hold, with the only changes $J=0$ and $R = \frac{|F \cos \varphi_3|}{|Z|}.$

What about the case $J=0$ and $\cos \varphi_3 = 0$? To get there by a limiting procedure of the $J\neq0$ equations, we need to be more careful. The condition $J=0$ implies $C_1(t)^2=C_2(t)^2,\quad \forall \,t \geq 0.$ For simplicity, we take initial conditions $C_1(0)=C_2(0) \neq 0$ so that $C_1(t) = C_2(t),\quad \forall \,t\geq 0.$ The condition $\cos \varphi_3=0$ along with the conservation law (\ref{eq:new_const}) then imply that $\sin \varphi(t) = 0, \quad \forall \,t\geq 0.$ For simplicity we take $\varphi(t) = 0, \quad \forall \,t\geq 0.$

Now, equation (\ref{def_R}) determines  $R \approx |J|/2 \to 0$ but equation (\ref{def_Delta}) gives a divergent shift: $\Delta \approx \log \left[4 C_1(0)^2/|J| \right] \to \infty.$  However a consistent limit is obtained for the physical quantities. For example, from equations (\ref{C_1(tau)_C_2(tau)_squared}) we obtain the amplitudes $C_1(t)^2 \approx C_2(t)^2 \approx C_1(0)^2 \exp(2\,Z\,\tau)\,.$ The evolution equation for $\tau$ is still of the form 
\begin{equation}
\label{J=0_bounded_eqs_for_tau} 
\left[\frac{\mbox{d}\tau}{\mbox{d}t}\right]^2 = E(0) -  V_{\lim}(\tau)\,,
\end{equation}
but with a new potential $V_{\lim}(\tau) = {\displaystyle \lim_{J\to0} V(\tau)},$ which reads 
\begin{equation}
 \label{potential_lim}
V_{\lim}(\tau) = C_1(0)^2  \exp(2\,Z\,\tau) - 2 F\,\tau\,\sin \varphi_3.
\end{equation}

Recall that $\cos \varphi_3 = 0\,.$ We conclude that the new potential $V_{\lim}(\tau)$ is attractive if and only if $\sin\varphi_3 = \mathrm{sgn}\left(Z\,F\right).$ This gives two separate cases, depending on the initial conditions: a bounded case and an unbounded case.

\subsubsection{The bounded sub-case of $J=0$, $\cos \varphi_3=0$}
\label{sec-bounded}

The bounded sub-case is obtained if we set $\sin\varphi_3 = \mathrm{sgn}\left(Z\,F\right).$ Although there is no simple solution of the system (\ref{J=0_bounded_eqs_for_tau}) in terms of elliptic functions, it is illustrative to obtain explicit expressions for the turning points, and consequently for the maximum and minimum values of the total energy (\ref{def_energy}), in terms of product log functions. The equations $V_{\lim}(\tau_{\min}) = V_{\lim}(\tau_{\max}) = E(0)$ have solutions
\begin{align*}
\tau_{\min/\max} &= -\frac{E(0)\sgn\left(Z\right)}{2 |F|} \\
&\quad - \frac{1}{2Z } W_k\left(-\frac{|Z| C_1(0)^2}{|F|}\exp\left(-\frac{E(0) |Z|}{|F|}\right)\right),
\end{align*}
where $k\in\mathbb{Z}$ denotes the different branches of the product log function $W_k(z)$, also known as the Lambert W-function. 
Notice that the argument of $W_k$ appearing above is strictly negative in this case, restricted to the interval $\left[-1/e,0\right],$ with the lower bound being obtained only when $C_3(0) = 0$. Since we require real solutions, we are limited in choice to the two branches $W_0$ and $W_{-1}$.

Substituting these expressions for the turning points into equation (\ref{eq_E}), we find the maximum and minimum values of the total energy of the system:
\begin{equation}
\label{energy_bounded}
E_{\min/\max} = -\frac{|F|}{|Z|} W_k\left(-\frac{|Z| C_1(0)^2}{|F|}\exp\left(-\frac{E(0) |Z|}{|F|}\right)\right),
\end{equation}
where we take $k=0$ for the minimum energy and $k=-1$ for the maximum.

\subsubsection{The unbounded sub-case of $J=0$, $\cos \varphi_3=0$}

\label{sec-unbounded}

The unbounded sub-case is obtained if we set $\sin\varphi_3 = -\mathrm{sgn}\left(Z\,F\right).$ In this case, the potential $V_{\lim}(\tau)$ in equation (\ref{potential_lim}) has a slope of fixed sign. Depending on the initial conditions, the motion can have one turning point but then the variable $Z\,\tau(t)$ will tend to $-\infty$ in the asymptotic way $Z\,\tau(t) \approx -|Z\,F|\, t^2$, so that the corresponding amplitudes' squares will behave as $C_3(t)^2 \approx 4\,F^2\,t^2\,, \quad C_1(t)^2 = C_2(t)^2 \approx K\,e^{-|Z\,F|\, t^2}\,,$ corresponding to unbounded growth of the forced mode and fast decay of the other two modes.

\section{Periods and Maximum amplitudes: parametric study}
\label{sec-parametricStudy}
Having established for $H=0$ the general integrability of
Eqs.~(\ref{reduced_system})  and the boundedness of the solution for almost all initial conditions (with the exception of one sub-case), we turn our attention to two important physical questions:
\begin{enumerate}
\item
 How does the physical period of oscillation depend on the parameters $Z, F$ and on the initial conditions $J$, $\varphi_3$, $E(0)$ of the problem? \\
\item How do the upper and lower bounds on the total energy of the system behave as a function of the parameters $Z$, $F$ and the initial conditions?\\
\end{enumerate}

\subsection{Summary of analytic results}
Before embarking on a numerical parametric study of the behaviour of the physical period and maximum energy, it is useful to try to answer the above questions in the few sub-cases where an explicit solution has been found:\\

\noindent {$\bullet$} In the case $H=0$, $\sin \varphi_3=0$ and $J \neq 0$, as detailed in Section \ref{sec-explicitSolution}, the full solutions for the amplitudes and phases are available explicitly. In particular, the period of oscillations is given by 
\begin{equation*}
T = \frac{2\,K(m)}{Z\,\sqrt{E(0)+R}},
\end{equation*}
where we have defined
\begin{equation*}
m  = \frac{E(0)-R}{E(0)+R} \in (0,1),\; R = \sqrt{\left(\frac{J}{2}\right)^2+\frac{{F\,}^2}{Z^2}},
 \end{equation*}
 and $K(m)$ is the complete elliptic integral of the first kind. 

In this case the total energy (\ref{def_energy}) of the system happens to be constant, equal to $E(0)$. This can be clearly identified in Figs. (\ref{fig-EmaxVsPhi3AndJ}) \& (\ref{fig-EmaxVsPhi3AndF}) where the surfaces for maximum and minimum energy coincide on the line given by $\sin\varphi_3=0$.\\\\
\noindent {$\bullet$} In the bounded sub-case of $H=0$, $\cos \varphi_3 = 0$ and $J = 0$, as detailed in Section \ref{sec-bounded}, although we do not have explicit expressions for the solutions for the amplitudes or even for the period of oscillations, we do have an exact expression for the maximum and minimum values of the total energy, from equation (\ref{energy_bounded}):
\begin{equation*}
E_{\min/\max} = -\frac{|F|}{|Z|} W_k\left(-\frac{|Z| C_1(0)^2}{|F|}\exp\left(-\frac{E(0) |Z|}{|F|}\right)\right),
\end{equation*}
where we take $k=0$ for the minimum energy and $k=-1$ for the maximum, and we have assumed for simplicity that $C_1(0)=C_2(0)$ and $\varphi=0.$\\

We now ask how these bounds obtained for the energy behave in the limit of arbitrarily large forcing. Recall that the Hamiltonian in the case $\cos \varphi_3 = 0$ reads
\begin{equation*}
H = 2ZC_1C_2C_3\sin\varphi.
\end{equation*}
Assuming each amplitude to be non-zero, fixing $H=0$ means that we have had to set $\varphi=0,$ irrespective of the value of $F.$ Therefore, by choosing appropriate initial amplitudes/phases so that $J=0,$ $\varphi_3=\pm\pi/2$ and $H=0,$ we can consider directly the limiting case when $F$ tends to infinity but with fixed initial conditions. Since $W_0$ is differentiable at zero, we have
\begin{equation*}
\lim_{F\rightarrow+\infty}E_{\min} = C_1(0)^2.
\end{equation*}
Calculating an equivalent expression for $E_{\max}$, is not quite so straightforward since $W_{-1}$ exhibits a singularity at $0$. However, knowing the asymptotic behaviour of $W_{-1}(x)$ as $x\rightarrow 0^{-}$ \cite{DLMF2011}, we find that to leading order as $F\rightarrow+\infty$, 
\begin{equation}
\label{energy_max_asympt}
E_{\max} \approx \frac{|F|}{|Z|}\ln\left(\frac{|F|}{|Z|\,C_1(0)^2}\right) + E(0).
\end{equation}

\subsection{Numerical study for arbitrary values of $J$ and $\varphi_3$}

\begin{figure}
\begin{center}
\includegraphics[width=8.0cm]{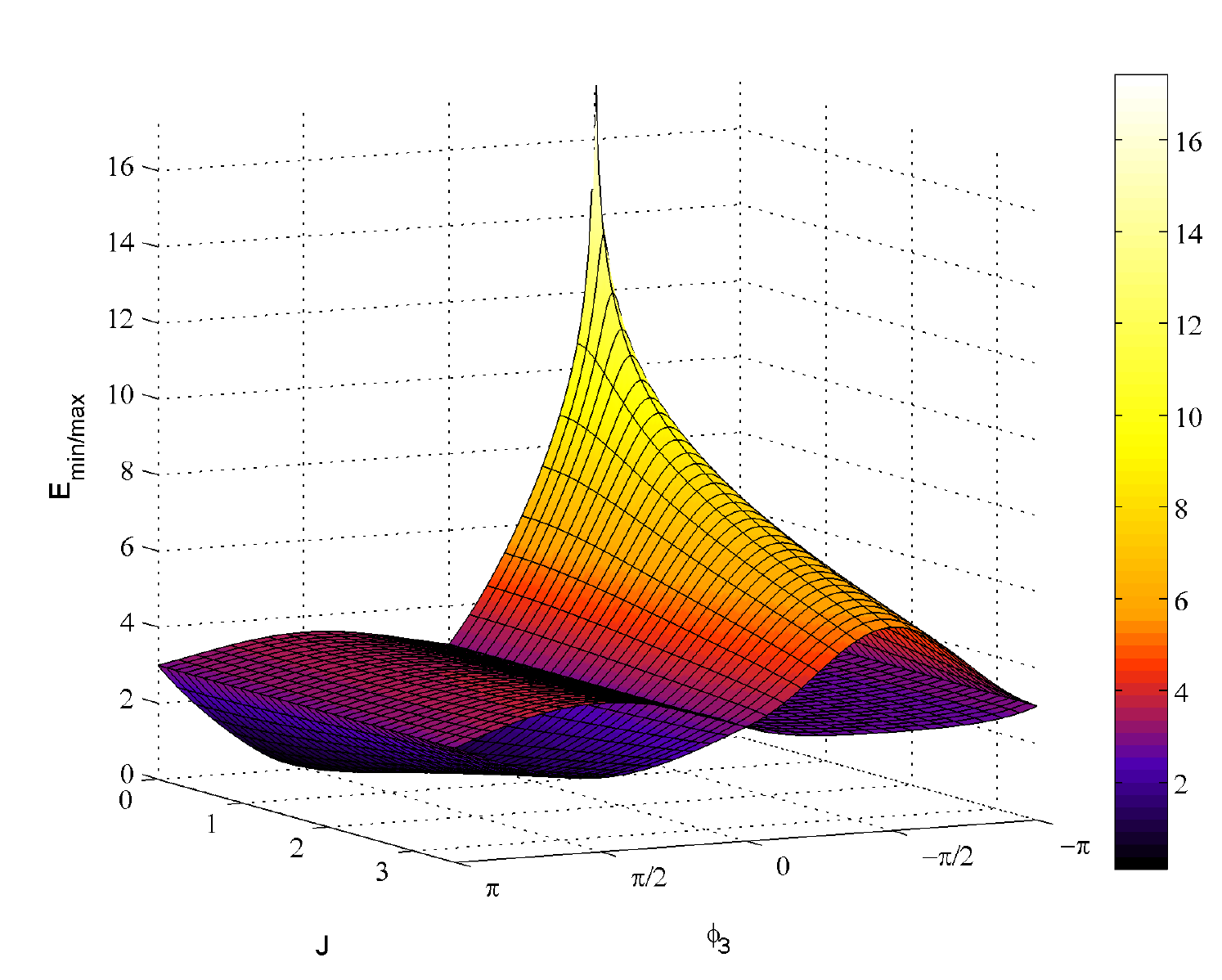}
\caption{\label{fig-EmaxVsPhi3AndJ} $E_\mathrm{max}$ and $E_\mathrm{min}$ as a function of $\varphi_3$ and $J$. As a choice of parameters and initial conditions, we have fixed $Z=1$, $E(0)=3$, $C_3(0)=1$ and $F=1$. Values of $C_1(0)$, $C_2(0)$ and $\varphi(0)$ can be found directly from the two conservation laws $J$ and $H=0$. }
\end{center}
\end{figure}

\begin{figure}
\begin{center}
\includegraphics[width=8.0cm]{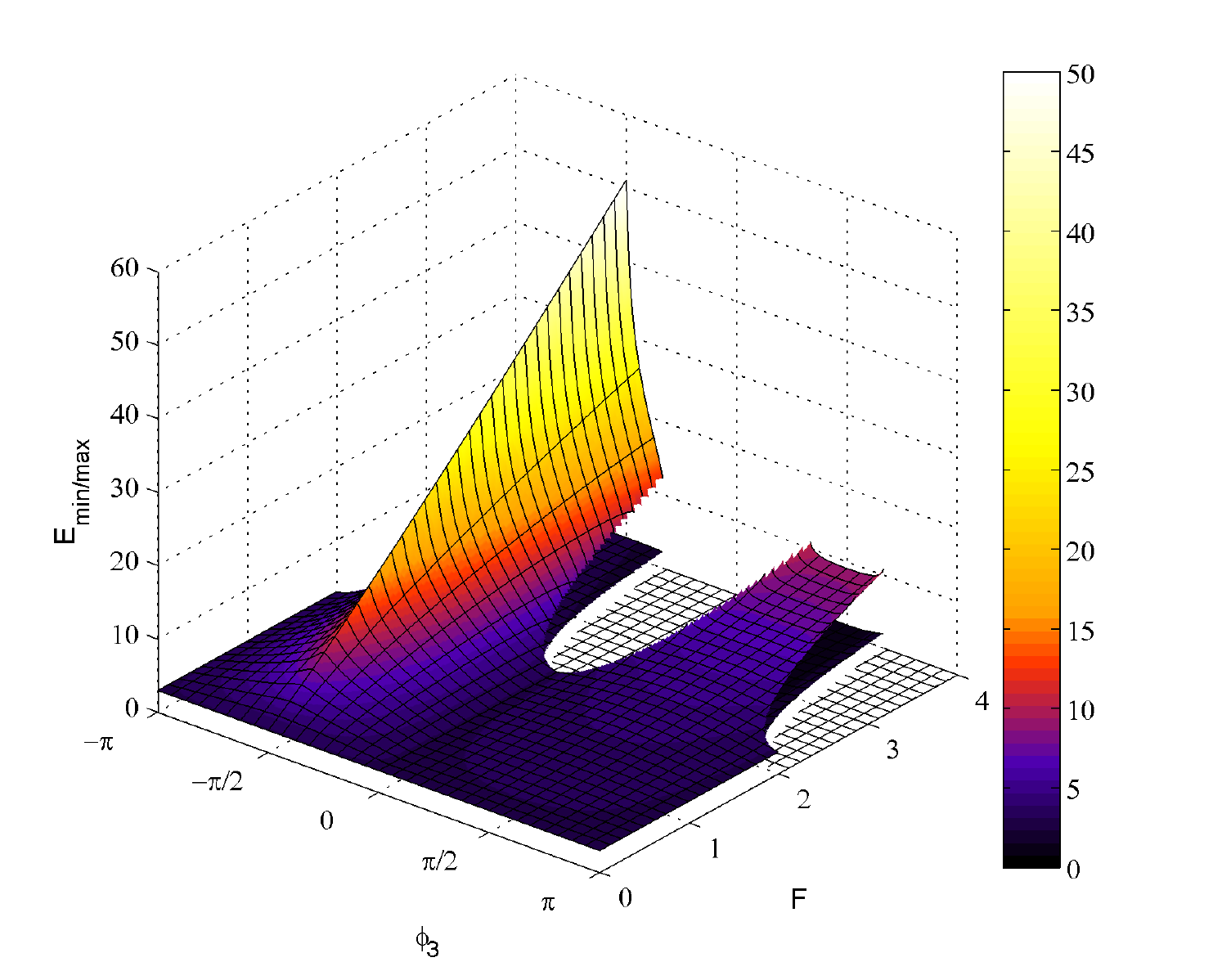}
\caption{\label{fig-EmaxVsPhi3AndF} $E_\mathrm{max}$ and $E_\mathrm{min}$ as a function of $\varphi_3$ and $F$. As a choice of parameters and initial conditions, we have fixed $Z=1$, $E(0)=3$, $C_3(0)=1$ and $J=0.1$. Values of $C_1(0)$, $C_2(0)$ and $\varphi(0)$ can be found directly from the two conservation laws $J$ and $H=0$.}
\end{center}
\end{figure}

\noindent \textbf{Maximum energy.} When neither of the two analytic cases detailed in the above bullet points apply, the two roots $\tau_{\min}$ and $\tau_{\max}$ must be computed numerically from equations (\ref{turning_points:xi}). The resulting values of $E_{\max}$ against varying $\varphi_3$ and $J$ under fixed forcing are shown in figure~\ref{fig-EmaxVsPhi3AndJ}. The unbounded sub-case as $J\rightarrow 0, \cos \varphi_3 = 0$ can clearly be identified. If instead we fix $J>0$ and allow the strength of the forcing $F$ to vary, we obtain figure~\ref{fig-EmaxVsPhi3AndF}. Here we observe a less-than-quadratic growth in the maximal energy in response to forcing, in a similar fashion as shown in the asymptotic result (\ref{energy_max_asympt}).\\

\begin{figure}
\begin{center}
\includegraphics[width=8.0cm]{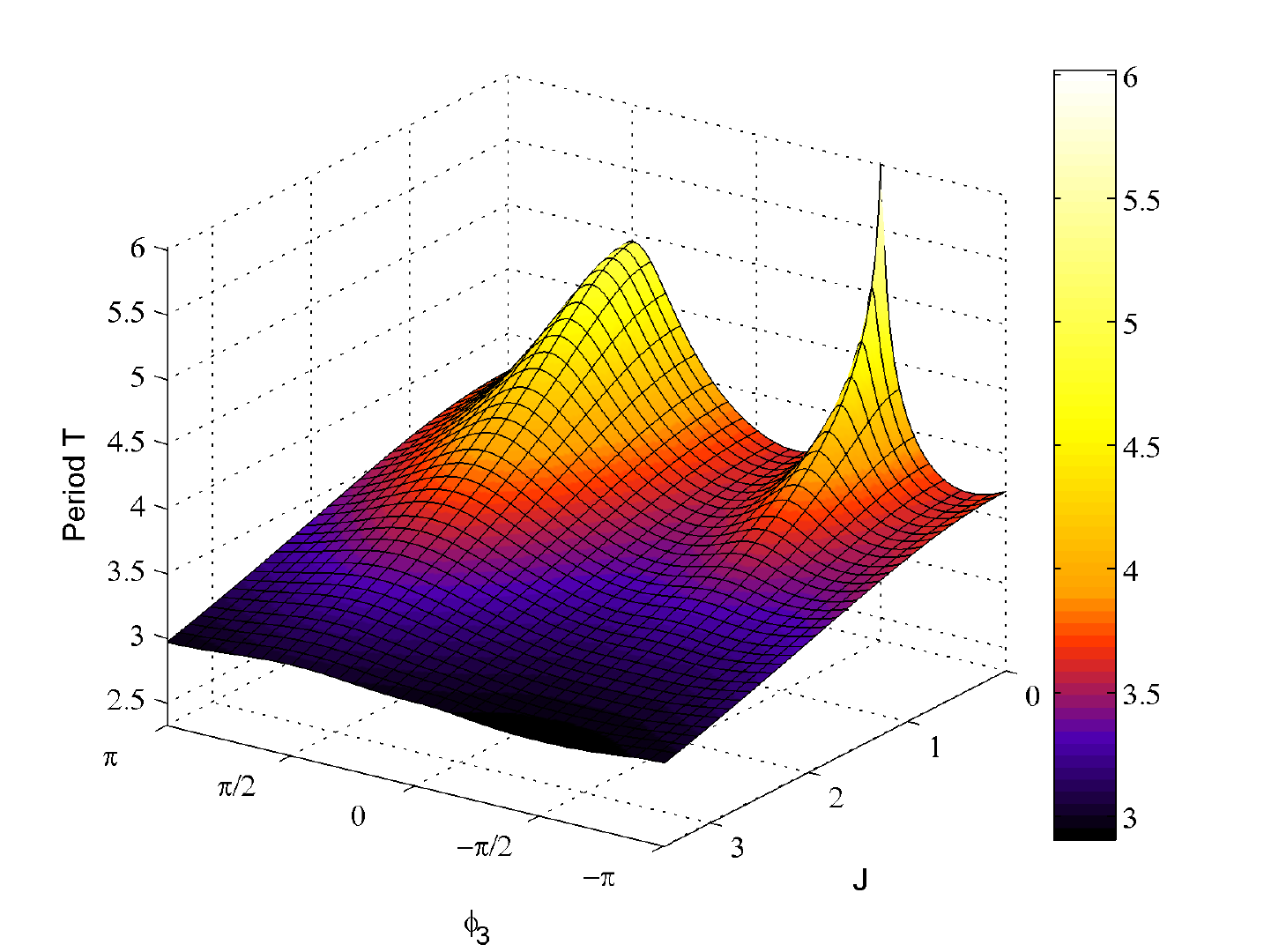}
\caption{\label{fig-periodVsPhi3AndJ} Period as a function of $\varphi_3$ and $J$. As a choice of parameters and initial conditions, we have fixed $Z=1$, $E(0)=3$, $C_3(0)=1$ and $F=1$. Values of $C_1(0)$, $C_2(0)$ and $\varphi(0)$ can be found directly from the two conservation laws $J$ and $H=0$. }
\end{center}
\end{figure}

\begin{figure}
\begin{center}
\includegraphics[width=8.0cm]{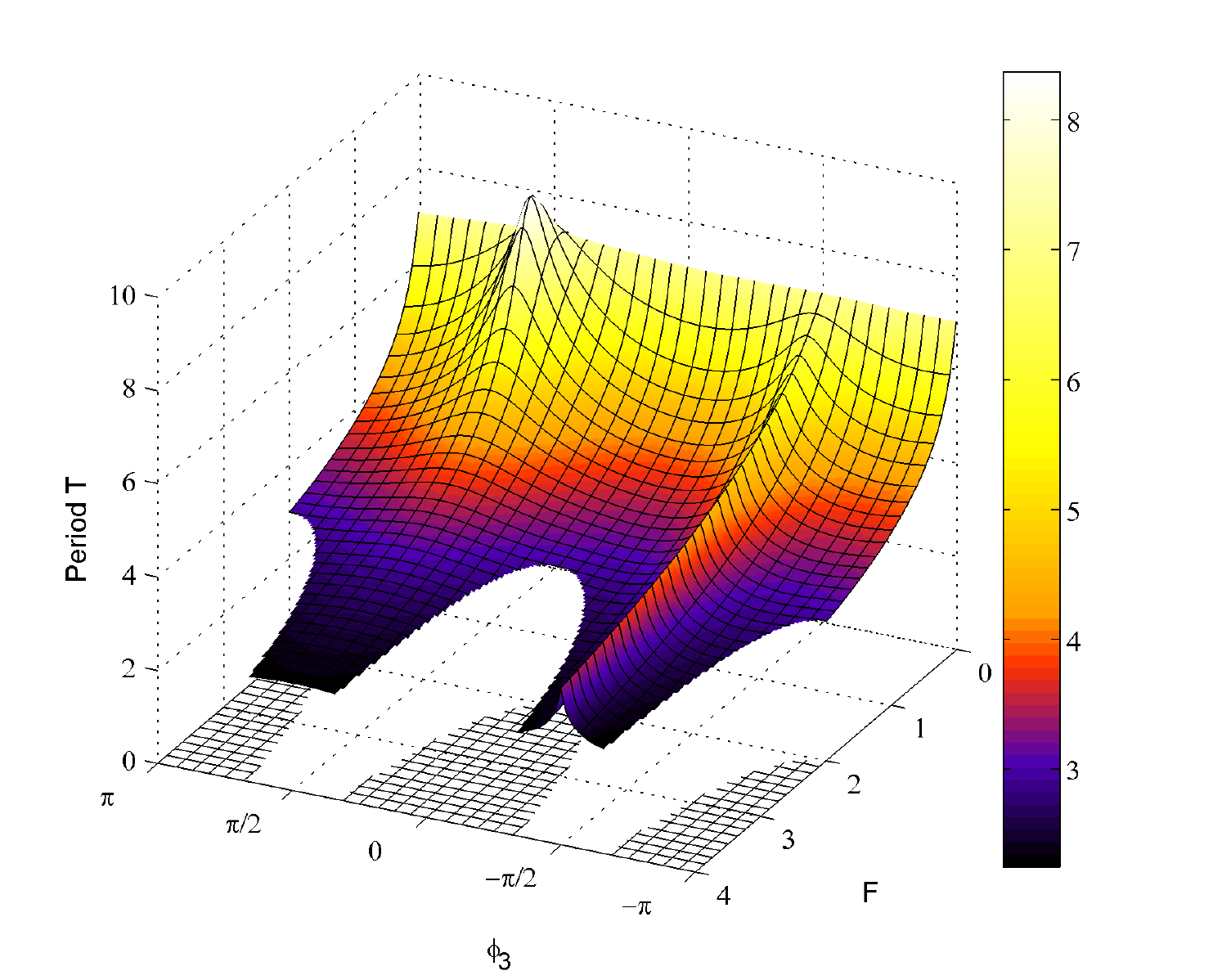}
\caption{\label{fig-periodVsPhi3AndF} Period as a function of $\varphi_3$ and $F$. As a choice of parameters and initial conditions, we have fixed $Z=1$, $E(0)=3$, $C_3(0)=1$ and $J=0.1$. Values of $C_1(0)$, $C_2(0)$ and $\varphi(0)$ can be found directly from the two conservation laws $J$ and $H=0$.}
\end{center}
\end{figure}

\noindent \textbf{Period of oscillations.} Since we only have an explicit expression for the period when $\sin \varphi_3 = 0,$ this too must be calculated numerically. After calculating numerically $\tau_{\max}$ and $\tau_{\min}$, integrating Eq. (\ref{differential_eqn_tau}) gives the period
\begin{equation*}
T = 2\int^{\tau_{\max}}_{\tau_{\min}} \frac{\diffd u}{\sqrt{E(0) - V(u)}}.
\end{equation*}
This integral can be evaluated numerically, with special care being taken where the integrand has singularities at the two end-points. The results are shown in figures \ref{fig-periodVsPhi3AndJ} and \ref{fig-periodVsPhi3AndF}.\\

One thing to note with regards to the figures, is that setting $H=0$ restricts the possible range of parameters that can be taken. The effect of this can be seen in figures~\ref{fig-EmaxVsPhi3AndF} and \ref{fig-periodVsPhi3AndF}, where forbidden parameter values are indicated by `hashed' regions at the base of the plot.

\section{Approximate Solutions in the integrable case $H=0$}
\label{sec-approximateSolutions}

As we have seen in Sec.~{\ref{sec-HZero}, in the integrable case $H=0$ only certain choices of parameter values lead to explicit formulae for the solution of equation (\ref{differential_eqn_tau}), in terms of elliptic functions. For other parameter values it is unlikely that such formulae exist and we have to resort to finding an approximate solution for $\tau(t)$. Since the potential $V(\tau) = R \cosh\left(2Z\tau+\Delta\right) -2 F\tau\sin\varphi_3$ is attractive, it is possible to approximate this potential by a polynomial in $\tau.$ A quartic polynomial is the best compromise between simplicity and accuracy: simplicity in that the resulting approximate evolution equation for $\tau(t)$ can be explicitly integrated, and accuracy in that one can capture nonlinear effects such as the dependence of the period on the parameters of the problem.

The problem is then reduced to compute the coefficients of this approximate potential. A natural place to start would be to obtain these coefficients directly from the fourth-order Taylor expansion of equation (\ref{differential_eqn_tau}), about the minimum of the potential $\tau=\tau_*.$ However, this `local' approach is doomed because it does not eliminate secular terms, and therefore the solution of the approximate equations differs more and more from the exact solution after each period.

Instead, the coefficients of this approximate potential must be chosen by a procedure that takes into account the \emph{global} character of the motion, i.e., the fact that the variable $\tau(t)$ can oscillate with high amplitude.
 
A quartic polynomial in $\tau$ has five free coefficients. We will determine two of them by requesting the approximated potential to have the same turning points as the exact potential. Thus, our approximate system is defined as follows:
$$\left[\frac{\mbox{d}\tau}{\mbox{d}t}\right]^2 = (\tau - \tau_{\min}) (\tau - \tau_{\max}) (\alpha_2\tau^2+\alpha_1\tau+\alpha_0)\,,
$$
where $\tau_{\min}, \tau_{\max}$ are the turning points, defined in equation (\ref{turning_points}), and $\alpha_0, \alpha_1, \alpha_2,$ are some unknown coefficients. This approximate system replaces the exact evolution equation (\ref{differential_eqn_tau}).

The remaining three coefficients will be determined by a minimisation procedure of the $L^2$ norm of the difference between the exact potential and the approximation.

\subsection{Approximate system, its solution and comparison with numerical simulations of the exact system}

In order to determine the unknown coefficients $\alpha_0, \alpha_1, \alpha_2,$ we impose a procedure that minimises the $L^2$ error, globally over the interval $[\tau_{\min}, \tau_{\max}].$ Let us define the function to be minimised:
\begin{equation*}
\Phi(\alpha_0,\alpha_1,\alpha_2) = \frac{1}{2} \int_{\tau_{\min}}^{\tau_{\max}}\left(P(\tau)-E(0)+V(\tau)\right)^2\;\diffd \tau,
\end{equation*}
where $P(\tau) \equiv (\tau - \tau_{\min}) (\tau - \tau_{\max}) (\alpha_2\tau^2+\alpha_1\tau+\alpha_0)$ and $V(\tau)$ is the potential energy, defined in equation (\ref{potential-bounded}). The function $\Phi$ is a positive-definite quadratic function of $(\alpha_0,\alpha_1,\alpha_2) \in \mathbb{R}^3.$ It has a single minimum, attained at the point $(\alpha_0,\alpha_1,\alpha_2)= (\alpha_0^*,\alpha_1^*,\alpha_2^*)\in \mathbb{R}^3.$ This point is the solution of the linear system
\begin{equation*}
\frac{\partial \Phi}{\partial\alpha_j}(\alpha_0^*,\alpha_1^*,\alpha_2^*) = 0\,, \quad j=0,1,2.
\end{equation*}
More explicitly, defining the real coefficients
\begin{eqnarray*}
\gamma_j &=& \int_{\tau_{\min}}^{\tau_{\max}} {\scriptstyle\tau^j  (\tau - \tau_{\min})^2 (\tau - \tau_{\max})^2 \;d \tau, \quad j =0, \ldots, 4},\\
\delta_j &=& \int_{\tau_{\min}}^{\tau_{\max}} {\scriptstyle\tau^j (\tau - \tau_{\min}) (\tau - \tau_{\max}) [E(0)-V(\tau)]\;d \tau, \quad j=0,1,2 }
\end{eqnarray*}
(elementary functions of $\tau_{\min}$ and $\tau_{\max}$), we obtain the system
\begin{equation*}
\left(
\begin{array}{ccc}
\gamma_0&\gamma_1&\gamma_2\\
\gamma_1&\gamma_2&\gamma_3\\
\gamma_2&\gamma_3&\gamma_4\\
\end{array}
\right)
\left(
\begin{array}{c}
\alpha_0^*\\
\alpha_1^*\\
\alpha_2^*
\end{array}
\right)
=
\left(
\begin{array}{c}
\delta_0\\
\delta_1\\
\delta_2
\end{array}
\right).
\end{equation*}

Since the matrix consisting of all the $\gamma$-coefficients is invertible as long as ${\tau_{\min}} < {\tau_{\max}}$, the solution for ${\alpha_j^*}$ is uniquely determined.

After calculating the coefficients $\alpha_j^*$ for $j=0,1,2$ as outlined above, we return to the approximate nonlinear differential equation
\begin{equation}
\label{approx_tau}
\left[\frac{\mbox{d}\tau}{\mbox{d}t}\right]^2 = (\tau - \tau_{\min}) (\tau - \tau_{\max}) (\alpha_2^*\tau^2+\alpha_1^*\tau+\alpha_0^*)\,.
\end{equation}

Solving this equation for $\tau(t)$ is now a matter of correctly evaluating an elliptic integral. To do this we must first reduce equation (\ref{approx_tau}) to Legendre normal form. The method closely follows the approach detailed in \cite{Abram72}: the quartic polynomial above is written as a product of two quadratics, $Q_1(\tau)\equiv \alpha_2^*\tau^2 + \alpha_1^*\tau +\alpha_0^*$ and $Q_2(\tau)\equiv(\tau-\tau_{\min})(\tau-\tau_{\max}).$ We look for values of $\lambda$ such that $Q_1-\lambda Q_2$ is a perfect square of a first-degree binomial in $\tau.$ To that end we require that 
\begin{equation*}
4(\alpha_2^*-\lambda)(\alpha_0^*-\lambda\tau_{\min}\tau_{\max}) - (\alpha_1^*+(\tau_{\min}+\tau_{\max})\lambda)^2=0.
\end{equation*}
Solving this quadratic equation for $\lambda$ we get
\begin{equation*}
\lambda_{\pm} = \frac{-\omega\pm\sqrt{\omega^2-(\tau_{\max}-\tau_{\min})^2((\alpha_1^*)^2-4\alpha_0^* \alpha_2^*)}}{(\tau_{\max}-\tau_{\min})^2},
\end{equation*}
where we have defined
\begin{equation*}
\omega=2\alpha_0^*+\alpha_1^*(\tau_{\max}+\tau_{\min})+2\alpha_2^* \tau_{\max}\tau_{\min}.
\end{equation*}
Having found $\lambda_+$, $\lambda_-$ we can then write
\begin{eqnarray*}
Q_1-\lambda_-Q_2 &=& (\alpha_2^*-\lambda_-)\left(\tau +\frac{\alpha_1^*+(\tau_{\max}+\tau_{\min})\lambda_-}{2(\alpha_2^*-\lambda_-)}\right)^2,\\
Q_1-\lambda_+Q_2 &=& (\alpha_2^*-\lambda_+)\left(\tau +\frac{\alpha_1^*+(\tau_{\max}+\tau_{\min})\lambda_+}{2(\alpha_2^*-\lambda_+)}\right)^2.
\end{eqnarray*}
Solving this system for $Q_1$ and $Q_2$ we obtain
\begin{eqnarray*}
Q_1 &=& \frac{\lambda_+(\alpha_2^*-\lambda_-)}{\lambda_+-\lambda_-}(\tau+\alpha)^2 -\frac{\lambda_-(\alpha_2^*-\lambda_+)}{\lambda_+-\lambda_-}(\tau+\beta)^2,\\
Q_2 &=& \frac{\alpha_2^*-\lambda_-}{\lambda_+-\lambda_-}(\tau+\alpha)^2 -\frac{\alpha_2^*-\lambda_+}{\lambda_+-\lambda_-}(\tau+\beta)^2,
\end{eqnarray*}
where we have defined the two constants
\begin{eqnarray*}
\alpha  &=& \frac{\alpha_1^*+(\tau_{\max}+\tau_{\min})\lambda_-}{2(\alpha_2^*-\lambda_-)},\\
\beta &=& \frac{\alpha_1^*+(\tau_{\max}+\tau_{\min})\lambda_+}{2(\alpha_2^*-\lambda_+)} .
\end{eqnarray*}

The method will work if and only if $\lambda_{\pm}$ are real and different. Since the coefficients $\alpha_j^*$, $\omega$, $\tau_{\max}$, $\tau_{\min}$ are all real, the condition for the method to work is equivalent to the condition $\omega^2-(\tau_{\max}-\tau_{\min})^2((\alpha_1^*)^2-4\alpha_0^* \alpha_2^*)>0.$ In turn, this inequality can be restated in terms of the zeroes $\tau_+$, $\tau_-$ of $Q_1$ and $\tau_{\max}$, $\tau_{\min}$ of $Q_2,$ as follows: $(\tau_{\max}-\tau_+)(\tau_{\max}-\tau_-)(\tau_{\min}-\tau_+)(\tau_{\min}-\tau_-) > 0.$ Graphically, the latter inequality  is equivalent to the statement that the zeroes of $Q_1$ are either: (i) complex (which necessarily come in conjugate pairs), (ii) real, but the interval $(\tau_{\min},\tau_{\max})$ contains either both zeroes of $Q_1$ or none of them.

Can we establish that the method will always work for physically admissible values of the parameters $Z$, $F$ and initial conditions $R$, $\varphi_3$, $E(0)$? Let us assume that $\tau_+$, $\tau_-,$ the zeroes of $Q_1,$ are real. By looking at the minimisation problem, notice that the potential $V(\tau)$ is convex. Therefore, if any of the zeroes $\tau_+$, $\tau_-$ were in $(\tau_{\max}$, $\tau_{\min}),$ then the approximating polynomial $P(\tau) = (\tau - \tau_{\min}) (\tau - \tau_{\max}) Q_1(\tau)$ would have a zero inside the interval $(\tau_{\max},\tau_{\min}),$ and so 
$P(\tau)$ would not be concave on $[\tau_{\min},\tau_{\max}].$ It follows that a deformation of the coefficients $\alpha_j$ could be found so that $\Phi(\alpha_0,\alpha_1,\alpha_2)$ would decrease, thus violating the minimum principle that generated the point $\alpha_j^*.$

In practice, in numerical applications and for a wide range of parameters we have found that the zeroes of $Q_1$ are complex (and therefore, conjugate pairs). In this case the method simplifies substantially because the quantities $\lambda_{\pm}$ satisfy $\lambda_- < 0 < \lambda_+.$ 

So let us assume from here on, and for simplicity, that $\lambda_- < 0 < \lambda_+.$ If we now make the substitution $r = (\tau+\alpha)/(\tau+\beta)$, we can reduce equation (\ref{approx_tau}) to the normal form
\begin{equation*}
\left[\frac{d r}{d t}\right]^2 =\frac{\lambda_+ (\alpha_2^*-\lambda_-)^2}{(\lambda_+-\lambda_-)^2}(\beta-\alpha)^2(a^2+r^2)(r^2-b^2),
\end{equation*}
where $a,b$ are positive parameters defined by
\begin{eqnarray*}
a^2 &=& \max \left\{-\frac{(\alpha_2^*-\lambda_+)}{(\alpha_2^*-\lambda_-)},-\frac{\lambda_-(\alpha_2^*-\lambda_+)}{\lambda_+(\alpha_2^*-\lambda_-)}\right\}, \\
b^2 &=& - \min \left\{-\frac{(\alpha_2^*-\lambda_+)}{(\alpha_2^*-\lambda_-)},-\frac{\lambda_-(\alpha_2^*-\lambda_+)}{\lambda_+(\alpha_2^*-\lambda_-)}\right\}.
\end{eqnarray*}
For physically sensible solutions (i.e., $\tau(t) \in \mathbb{R}$) we must choose the branch $r(t)^2 > b^2$ for all $t \geq 0$. We obtain, in terms of the Jacobi elliptic function,
\begin{equation*}
r(t) = b\;\mbox{nc}\left(\left. \frac{4K(m_{0})}{T_{0}}t+u_0\right|m_{0}\right),
\end{equation*}
defined by the physical period,
\begin{equation*}
T_{0} = \frac{2K(m_{0})}{|Z||\beta-\alpha|\left[ \frac{\lambda_+ \alpha_1^*(\alpha_2^*-\lambda_-)^2}{(\lambda_+-\lambda_-)^2}(a^2+b^2)\right]^{\frac{1}{2}}},
\end{equation*}
and elliptic modulus
\begin{equation*}
m_{0} = \frac{a^2}{a^2+b^2}.
\end{equation*}
The value of the shift $u_0$ can be found directly in terms of the incomplete elliptic integral of the first kind: 
\begin{equation*}
u_0 = s_1 F\left(\left.\arccos\left(\frac{b(\Delta+\beta)}{\Delta+\alpha}\right)\right|m_{0}\right),
\end{equation*}
with $s_1\in\{-1,+1\}$ being given by
\begin{equation*}
s_1 = \sgn\left(ZC_3(0)(\beta-\alpha)\right).
\end{equation*}
Undoing the transformations applied to $\tau$ we finally uncover the solution
\begin{equation}
\label{solu_approx}
\tau(t) = \frac{b\beta-\alpha\;\mbox{cn}\left(\left. \frac{4K(m_{0})}{T_{0}}t+u_0\right|m_{0}\right)}{\mbox{cn}\left(\left. \frac{4K(m_{0})}{T_{0}}t+u_0\right|m_{0}\right)-b}\,.
\end{equation}
Approximate expressions for the squares of the amplitudes $C_j(t)^2$ are obtained directly from equations (\ref{C_1(tau)_C_2(tau)_squared}) and (\ref{C_3_squared-tau}), and the dynamical phase is obtained from equation (\ref{dynamical_phase_H=0}). Plugging the solution (\ref{solu_approx}) of our approximate system, into these expressions, we generate plots that can be compared with the numerical solutions of the original system (\ref{reduced_system_H_is_zero-new_cons}). The results, shown in figures \ref{fig_C1_approx_L2} to \ref{fig_phase_approx_L2}, show excellent agreement. We have validated numerically, for a wide range of choices of parameters, the following condition of boundedness of the approximate solution (\ref{solu_approx}): $b>1$ or, equivalently, $\alpha_2^*<0.$

\begin{figure}
\begin{center}
\subfigure[ $C_3(t)^2$ with $F=1$]{
\includegraphics[width=40mm]{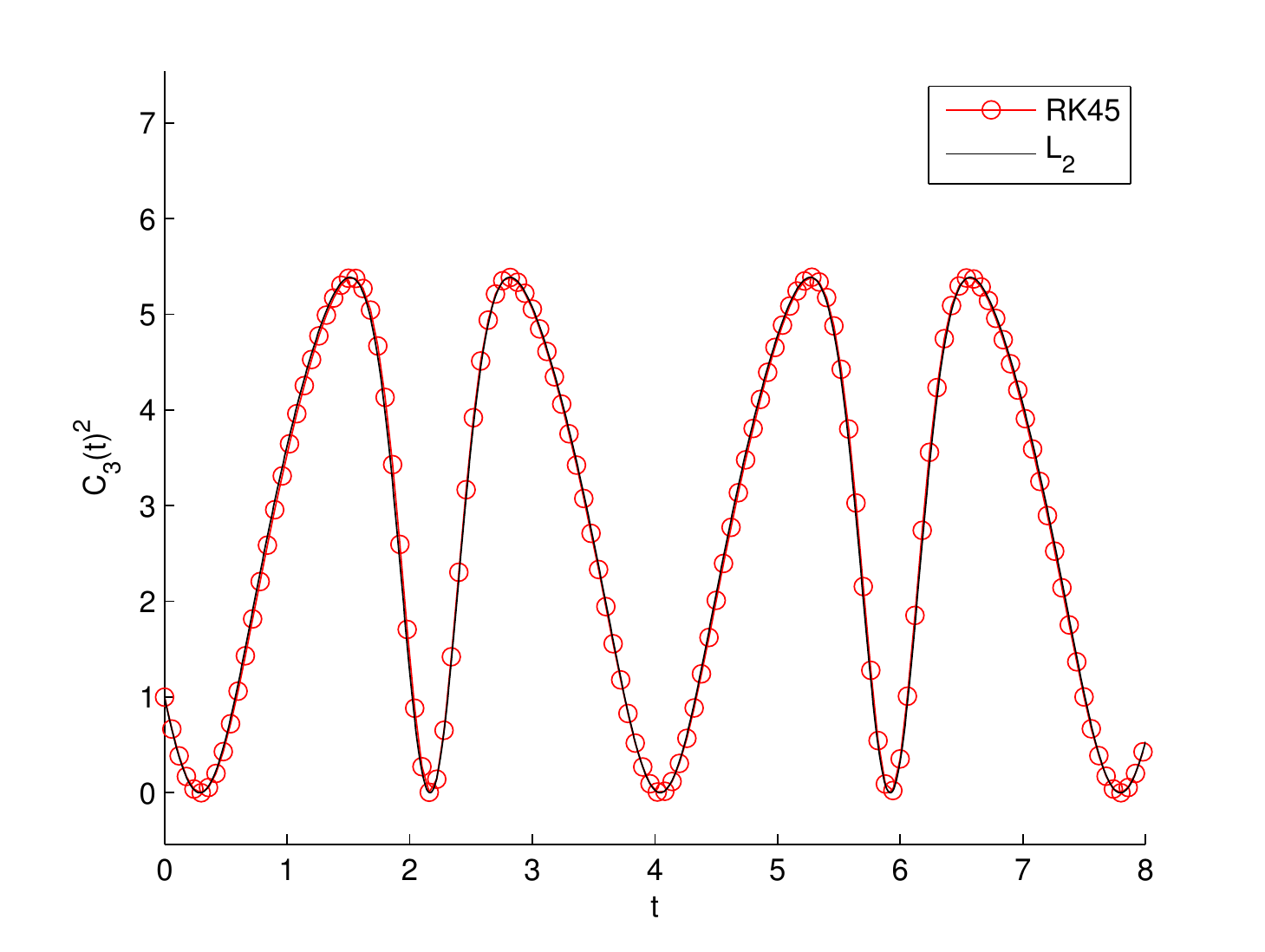}
\label{fig_C1_approx_L2}
}
\subfigure[ $C_3(t)^2$ with $F=10$]{
\includegraphics[width=40mm]{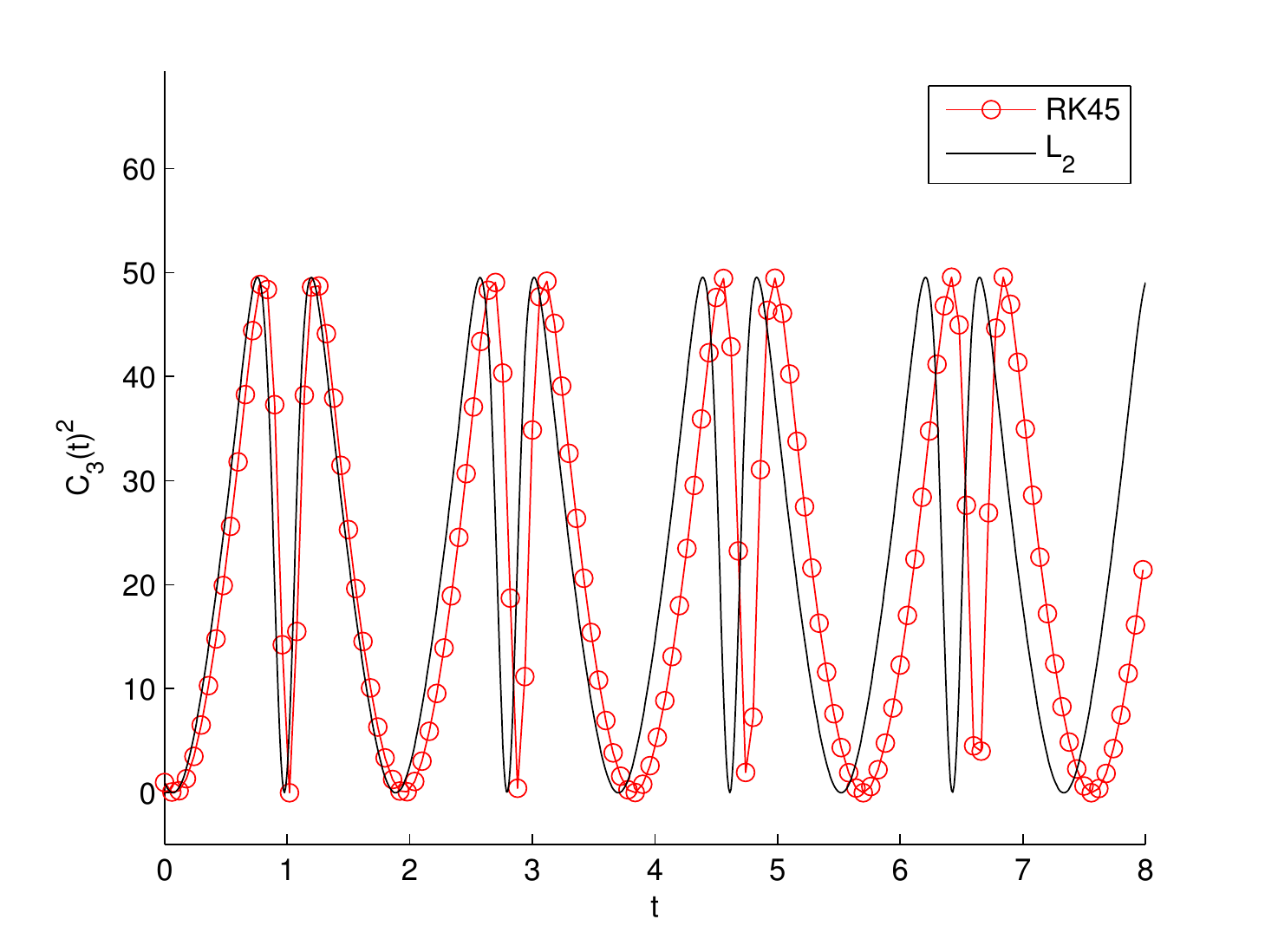}
\label{fig_C2_approx_L2}
}
\subfigure[ $\varphi(t)$ with $F=1$]{
\includegraphics[width=40mm]{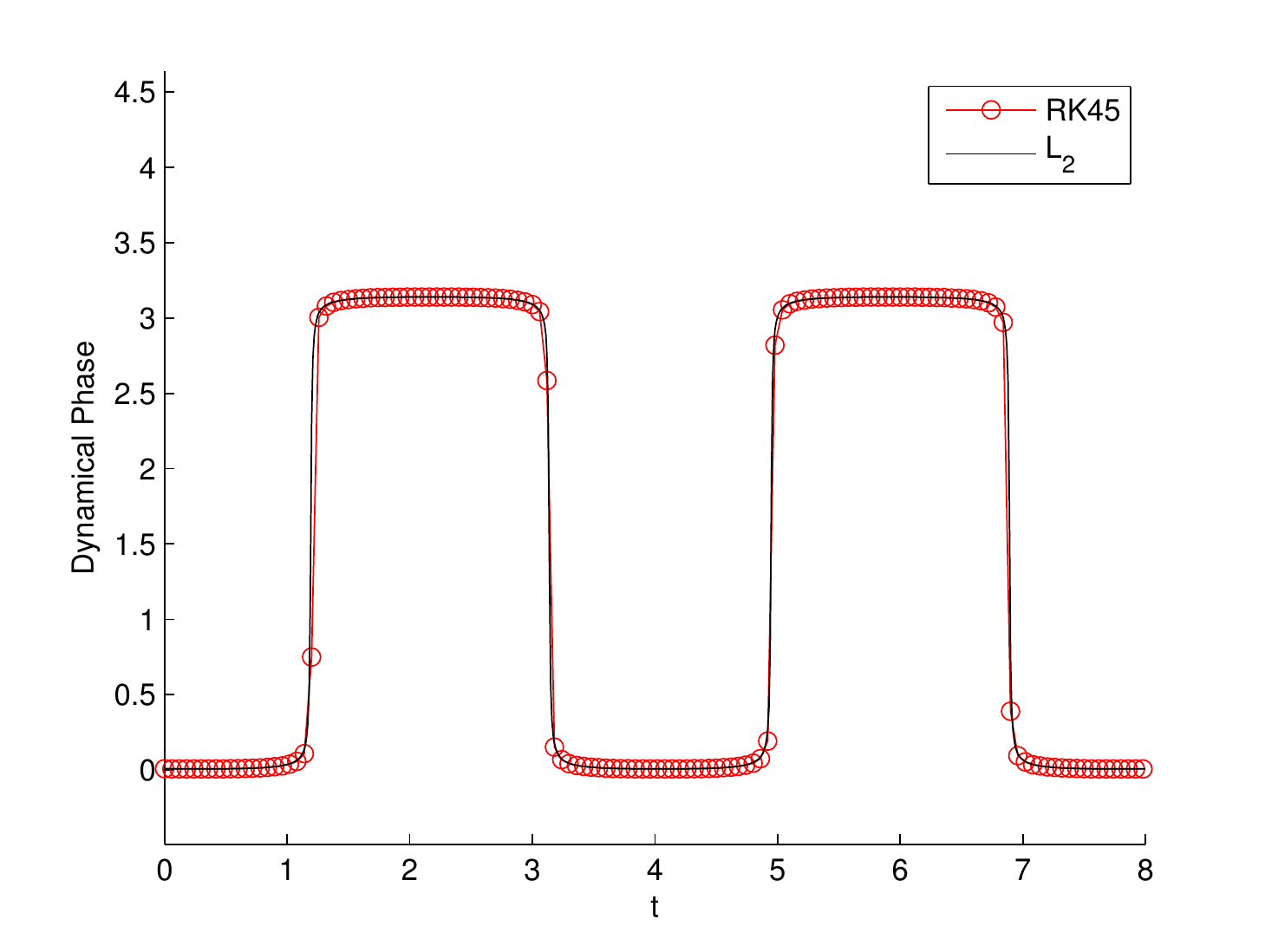}
\label{fig_C3_approx_L2}
}
\subfigure[ $\varphi(t)$ with $F=10$]{
\includegraphics[width=40mm]{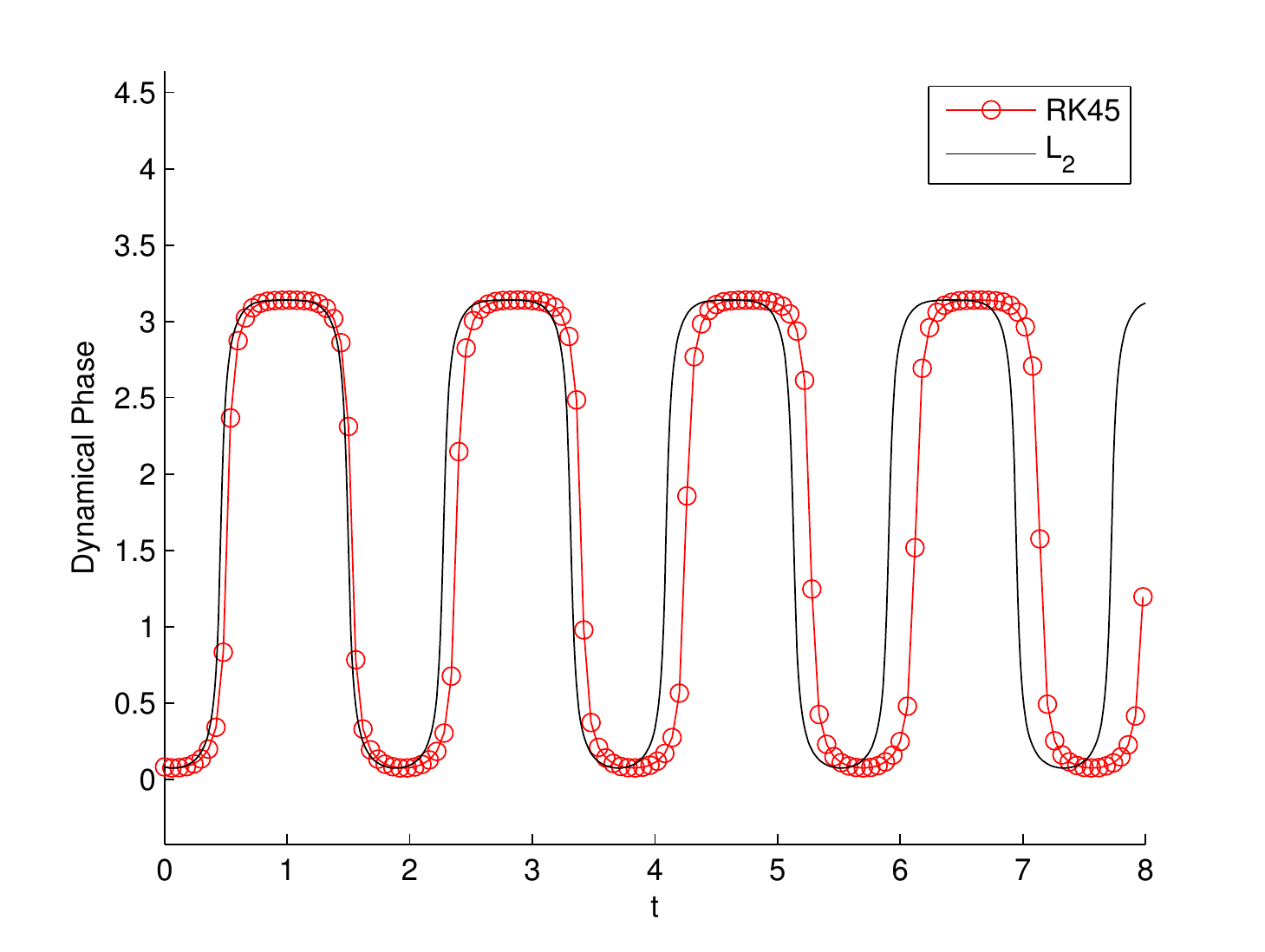}
\label{fig_phase_approx_L2}
}
\caption{Comparison of the $L^2$-method approximate solutions for the square amplitudes and dynamical phase. The corresponding numerically generated Runge-Kutta solutions are also included as a direct comparison against different forcing strengths. Choices for parameters: $\varphi_3(0) = -0.99\frac{\pi}{2}$, $E(0)=3$, $J=1$, $Z=1$ and $\varphi(0)$ determined by setting $H=0$. }
\end{center}
\end{figure}

\section{Poincar\'{e} sections and evidence for boundedness when $H\neq 0$}
\label{sec-HNonZero}
When the Hamiltonian is non-zero, we have no \emph{a priori} guarantee over integrability of 
the forced triad system, due to the system's formal equivalence to an autonomous volume-preserving 3-dimensional system of first order equations. If we try to reduce the forced triad system to an explicit 3-dimensional system we face yet another difficulty: the generic appearance of coordinate singularities, as exemplified in equations (\ref{p-a_equations_of_motion}) at $C_1 C_2 = 0.$ 

Fortunately, we can solve this difficulty completely and simplify matters considerably through an intelligent choice of variables. The procedure works only when $H \neq 0$, and is based on the observation that $\varphi_3$ is monotonic and so it becomes a proxy for time. Let us define the new variables,
\begin{equation*}
p = \Re\{ B_1B_2\},\;q = \Im\{ B_1B_2\},\\
\end{equation*}
with $\varphi_3 = \mbox{Arg}\;B_3$ as before. Using the conservation law $J$, we observe that 
\begin{equation*}
\left(|B_1|^2 + |B_2|^2\right)^2 - J^2 = 4|B_1|^2|B_2|^2,
\end{equation*}
and rewriting the right-hand term $|B_1|^2|B_2|^2 = p^2 + q^2$, we derive the remarkable formula
\begin{equation*}
|B_1|^2 + |B_2|^2 = \sqrt{J^2 + 4(p^2+q^2)}.
\end{equation*}
Differentiating the new variables $(p,q,\varphi_3)$ with respect to time we obtain, after straightforward computations, the following three-dimensional system:
\begin{equation*}
\left\{
\begin{aligned}
\dot{p}&= Z\sqrt{J^2 + 4(p^2+q^2)}|B_3|\cos\varphi_3,\\
\dot{q}&= Z\sqrt{J^2 + 4(p^2+q^2)}|B_3|\sin\varphi_3,\\
\dot{\varphi_3}&=-\frac{H}{2|B_3|^2},
\end{aligned}
\right.
\end{equation*}
where the amplitude $|B_3|$ is a simple function of $(p,q,\varphi_3)$ and the Hamiltonian $H,$ which now reads
\begin{equation}
\label{H_pq}
H = 2Z|B_3|\left[\left(q-\frac{F}{Z}\right)\cos\varphi_3 - p\sin\varphi_3\right].
\end{equation}

In this representation, we see that the angle $\varphi_3$ is not only monotonically increasing or decreasing depending on $H$, but can also be read directly as the slope of the parametric curve $\left(p(t),q(t)\right)$. This follows simply from the relation
\begin{equation*}
\frac{\diffd q}{\diffd p} = \tan\varphi_3.
\end{equation*}
Furthermore, we see from this new representation of $H$ given in Eq.(\ref{H_pq}), that the condition for $|B_3|$ to remain finite is equivalent to the statement that the cross-product between the tangent vector, $\left(\dot{p},\dot{q}\right)$ and the off-centre radius vector $\left(p,q-F/Z\right)$ must remain non-zero. In essence, $\left(\dot{p},\dot{q}\right)\times\left(p,q-F/Z\right)$ must point out of the plane.

Using this equivalence it makes sense to shift the origin of $(p,q)$ coordinates and define a new pair of variables $(r,\phi)$ such that $p = r\cos\phi$ and $q = F/Z + r\sin\phi,$ or
\begin{eqnarray*}
r^2 &=&\left(q-F/Z\right)^2+p^2,\\
\tan\phi&=&\frac{q-F/Z}{p}.
\end{eqnarray*}
Differentiating these expressions with respect to time gives the following system:
\begin{equation*}
\left\{
\begin{aligned}
\dot{r} &= \frac{H}{2r}\sqrt{J^2+4(p^2+q^2)}\cot\left(\phi-\varphi_3\right),\\
\dot{\phi} &= -\frac{H}{2r^2}\sqrt{J^2+4(p^2+q^2)},\\
\dot{\varphi_3} &= -\frac{2Z^2r^2}{H}\sin^2\left(\phi-\varphi_3\right).
\end{aligned}
\right.
\end{equation*}
The Hamiltonian now reads
\begin{equation*}
H = 2Z|B_3|r\sin\left(\phi-\varphi_3\right).
\end{equation*}
We immediately see that $\phi$ as well as $\varphi_3$ must be monotonic, increasing or decreasing  identically with $\varphi_3$. Since we have defined $r$ to be positive, we also see that since we must necessarily have $|B_3|>0$ then either $\sin\left(\phi-\varphi_3\right) >0$ when $ZH>0$, or, $\sin\left(\phi-\varphi_3\right) <0$ when $ZH<0$. Depending on the choice of initial conditions, we therefore obtain the bounds
\begin{equation*}
\begin{cases}
0<\phi-\varphi_3<\pi, & \mbox{if}\; ZH>0;\\
-\pi<\phi-\varphi_3<0, & \mbox{if}\; ZH<0.
\end{cases}
\end{equation*}
These bounds cannot be saturated, otherwise we violate the assumption $H\not=0$.  The variable $\phi$ is monotonic, so we will define it as our new `time'. Introducing the variable $\theta\equiv\phi-\varphi_3$ we get the new system
\begin{equation*}
\left\{
\begin{aligned}
\frac{\diffd r}{\diffd\phi} &=-r\cot\theta,\\
\frac{\diffd\theta}{\diffd\phi} &= 1 - \frac{4Z^2r^4\sin^2\theta}{H^2\sqrt{J^2+4(p^2+q^2)}},
\end{aligned}
\right.
\end{equation*}
which is non-autonomous, but one where the `time'-dependence appears only in the $J^2 +4(p^2+q^2)$ term. Writing this term in these latest variables reads
\begin{equation*}
J^2 + 4(p^2+q^2) = 4\left[\left(r+\frac{F\sin\phi}{Z}\right)^2 + c(\phi)^2\right],
\end{equation*}
where we have defined the $\phi$-dependent, positive function $c(\phi)$ according to
\begin{equation*}
c(\phi)^2 = \frac{J^2}{4} + \frac{F^2\cos^2\phi}{Z^2}.
\end{equation*}

\noindent \textbf{One-dimensional particle interpretation.} One may ask what is gained by these successive transformations. After all, 
this newest version of the system appears to offer little improvement over 
the original one written in terms of the variables $p$ and $q$. Surprisingly, 
however, this new system has a novel one-dimensional particle interpretation 
similar to that obtained in the $H\not=0$ case described earlier. If we make 
one final transformation $h(\phi) = r(\phi)^{-1}$, we derive the 
one-dimensional particle representation:
\begin{equation}
\label{H_nonzero_oscillator}
\frac{\diffd^2h}{\diffd\phi^2} + h = \frac{2Z^2}{H^2h^3\sqrt{\left(h^{-1} + \frac{F\sin\phi}{Z}\right)^2 + c(\phi)^2}}.
\end{equation}
This can be interpreted as a driven harmonic oscillator, with driving force that is $2\pi$-periodic in $\phi$ but also dependent on the position $h$. Alternatively, we can write the system in terms of a potential $V(h,\phi)$:
\begin{equation*}
\ddot{h} + V_h(h,\phi) = 0,
\end{equation*}
where the `dot' now denotes derivative with respect to $\phi$, and subscripts denote partial derivatives. Integrating Eq. (\ref{H_nonzero_oscillator}) with respect to $h$, we evaluate this potential to be of the form,
\begin{align}
\nonumber
V(h,\phi) &= \frac{h^2}{2}+\frac{2Z^2}{H^2}\sqrt{\left(h^{-1} + \frac{F\sin\phi}{Z}\right)^2 + c(\phi)^2}\\
\nonumber
&-\frac{2ZF\sin\phi}{H^2}\asinh\left(\frac{h^{-1} + \frac{F\sin\phi}{Z}}{c(\phi)}\right) + f(\phi),\\
&
\end{align}
where $f(\phi)$ is a yet to be determined function that is dependent on $\phi$ only. This potential is attractive for all values of $\phi$, from which it follows that the solution must at least be bounded for finite `time'.

\begin{figure}[htbp]
\begin{center}
\subfigure[$H = 0.1$]{
\includegraphics[width=90mm]{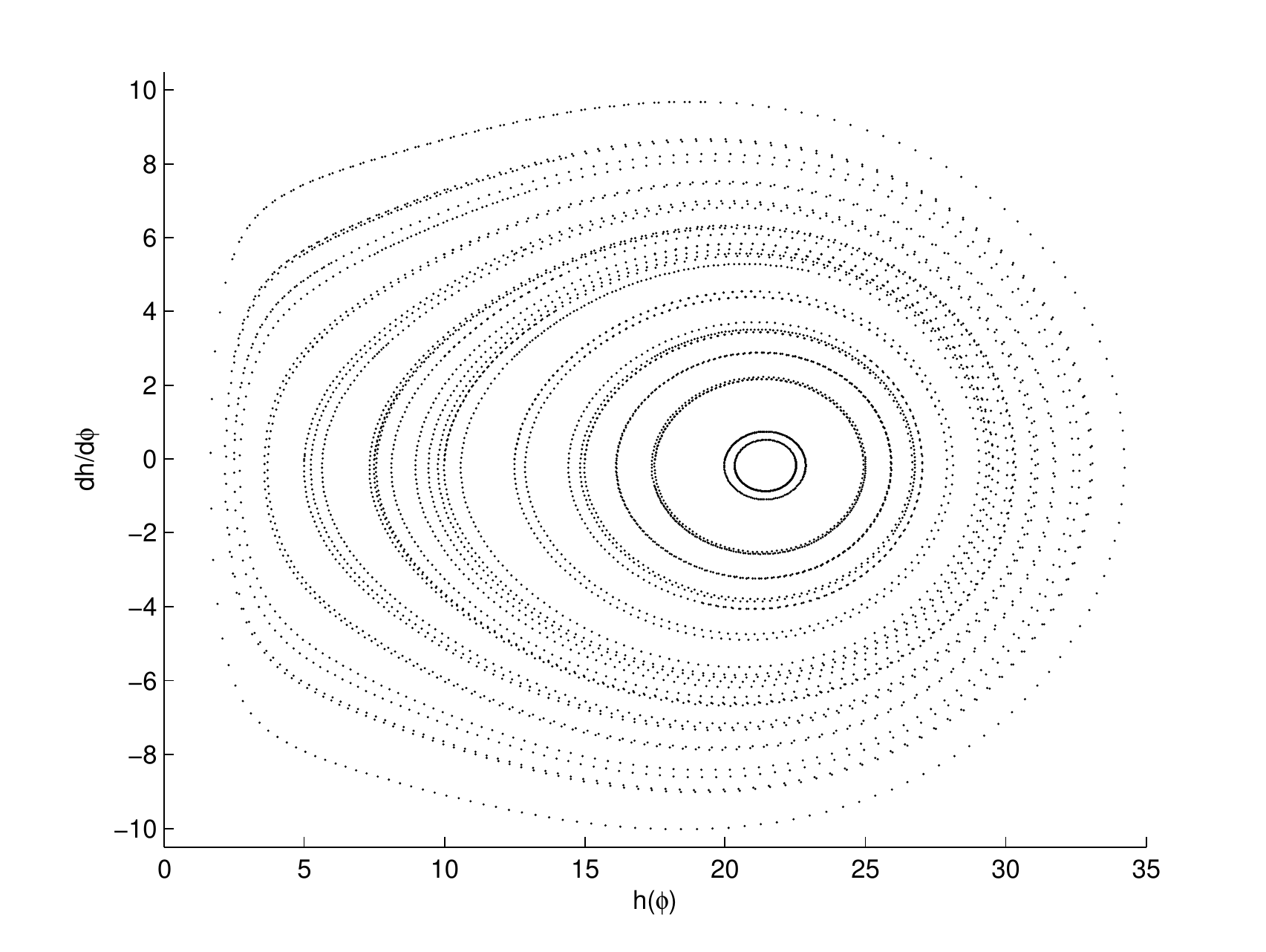}
\label{Poincare_H0p1}
}
\subfigure[ $H=1$]{
\includegraphics[width=90mm]{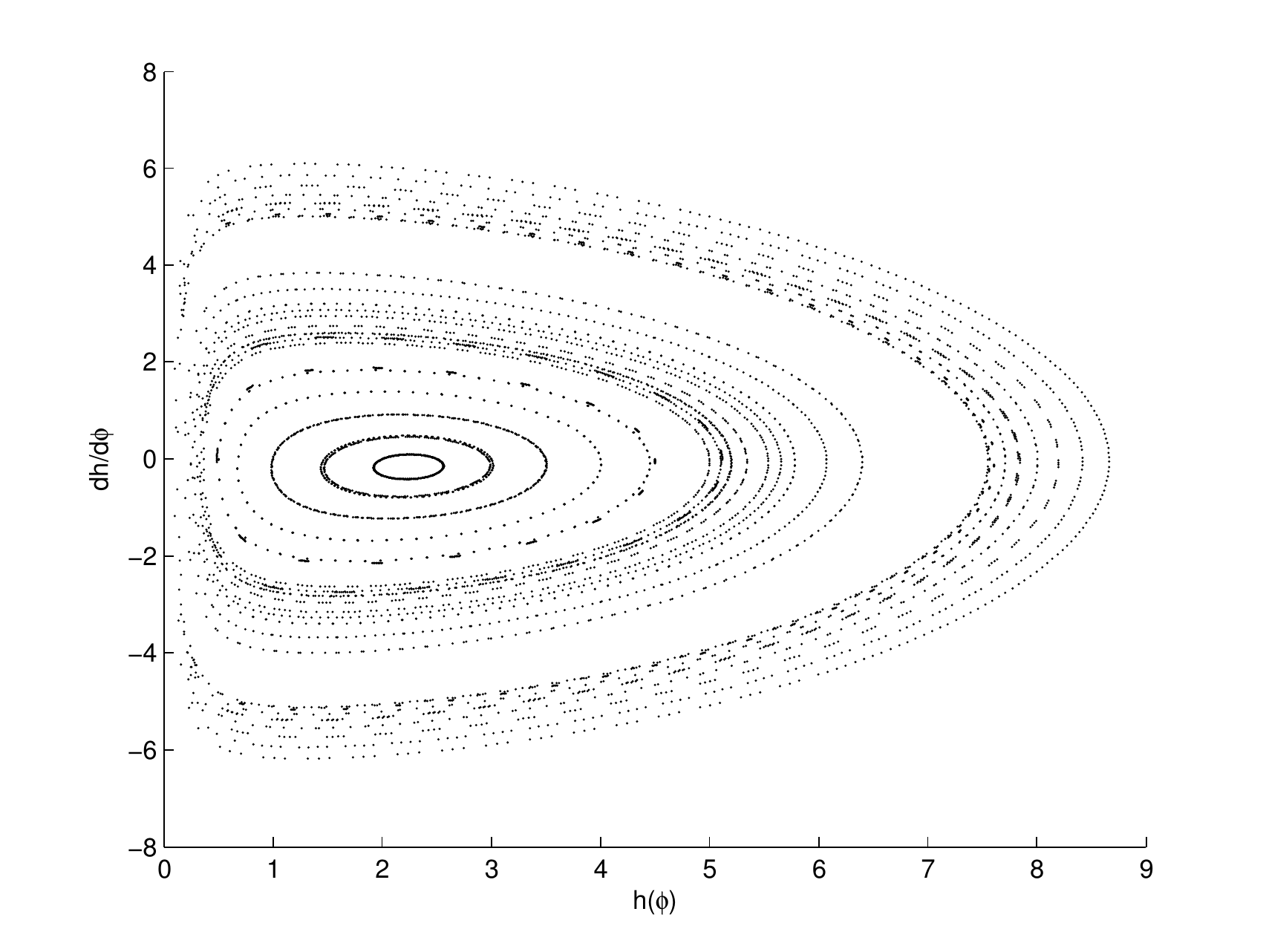}
\label{Poincare_H1}
}
\caption{\label{Poincare_sections}Poincar\'{e} sections produced for varying values of $H$. Here we have fixed the values of $Z=1$, $J=1$ and $F=1$ in both panes. Closed invariant curves can clearly be identified in each figure.}
\end{center}
\end{figure}

The important question that remains is whether the solution remains bounded for all time. Our analytical work is in progress, based on Levi's theory \cite{L1991}, and will be reported elsewhere. Our numerical studies suggest that the answer is positive. We present now the evidence. Due to the periodic nature of the forcing term, we define a Poincar\'{e} map, $P:S\rightarrow S$, where $S=\mathbb{R}_{>0}\times\mathbb{R}$ is the set of initial conditions of the form $(h(0),h_\phi(0)),$ and the transverse section is given by $\phi_n=2n\pi$ corresponding to $H<0$, and $\phi_n=-2n\pi$ if $H>0,$ taking $n\in\mathbb{N}_0$. Successive iterations of the map $P$ are then defined through the equation
\begin{equation*}
P^n(h(0),h_\phi(0)) = (h(\phi_n),h_\phi(\phi_n)),\;\,n\in\mathbb{N}_0.
\end{equation*}
Examples of some numerically generated Poincar\'{e} sections are given in figure \ref{Poincare_sections}. There is clearly no chaotic behaviour in the system, with the motion being quasi-periodic in nature, restricted to clearly visible closed invariant curves. Each of these curves is in fact a loop, suggesting that the motion remains bounded indefinitely. Furthermore, since the map is continuous when restricted to each of these one-dimensional loops, Brouwer's fixed-point theorem guarantees the existence of at least one fixed-point. It therefore follows that the system contains infinitely many periodic orbits, with period $2\pi$ in $\phi$, one of which can clearly be identified on the line $h_\phi=0$ where these loops converge to a point.

\section{Conclusions}

We have presented an in-depth analysis of the triad equations
driven by external forcing, Eqs.~{\ref{equations_of_motion}. We obtained a 
complete understanding of the dynamics when the Hamiltonian is zero by
using conservation laws to reduce Eqs.~{\ref{equations_of_motion} to
the one-dimensional motion of a particle in a time-independent potential. In 
this case, the dynamics are integrable and bounded for almost all 
initial conditions with a single 
exception provided by the special initial condition discussed in 
Sec.~\ref{sec-unbounded}. Along the way, we presented several new explicit
solutions, analytic formulae for the period and maximum energy of the motion 
and a novel scheme for approximating the dynamics when an explicit solution
seems beyond reach. Our results for the case $H\neq 0$ are more qualitative
and based on reducing the problem to the one-dimensional motion of a particle 
in a time-periodic potential. Poincar\'{e} sections of the dynamics provide
strong empirical evidence that the amplitudes remain bounded when $H\neq 0$
and are typically quasi-periodic although periodic orbits also exist. 
Broadly speaking, our results illustrate two important points which are
likely to have relevance beyond the particular problem studied in this
article. The first is that, as illustrated by our boundedness results, the 
addition of forcing to a nonlinear wave system does not necessarily provide a 
continual source of energy for the system. This should provide pause for
thought for researchers interested in performing numerical simulations of
wave turbulence (where a constant source of energy is often required in order 
to compare with theory) since such simulations are often performed by adding 
an external forcing term to a Hamiltonian wave equation. The second general 
lesson to  be drawn from this work is the importance of phases in understanding 
the dynamics of resonantly interacting nonlinear waves. Since the dynamics
of phases are often more difficult to obtain theoretically, they are often
somewhat under-emphasized. Our results present some of the most striking 
examples yet of how varying an initial phase can result in completely 
different behaviour.

\section{Acknowledgements}
We acknowledge useful scientific discussions with P. Lynch and S. Nazarenko.
MDB acknowledges UCD support under projects SF304 and SF564.
CC acknowledges the support of the Engineering and Physical Sciences Research 
Council under grant No. EP/H051295/1.

\ifthenelse{\boolean{els}}
{
\section*{References}
\bibliographystyle{model1-num-names}
}
{
}


%
\end{document}